\documentclass[a4paper,12pt]{article}
\usepackage{amsmath,amssymb}
\usepackage{hyperref}
\usepackage{graphicx}
\usepackage{mathrsfs}
\usepackage{float}
\usepackage{ulem}
\hypersetup{
    colorlinks,%
    citecolor=blue,%
    filecolor=blue,%
    linkcolor=blue,%
   urlcolor=blue,
   linktoc=page
}
\usepackage{chngcntr}
\counterwithin{equation}{section}

\newcommand{\be}{\begin{equation}}
\newcommand{\ee}{\end{equation}}
\newcommand{\beq}{\begin{equation}}
\newcommand{\eeq}{\end{equation}}
\newcommand{\f}{\frac}
\newcommand{\s}{\sqrt}
\newcommand{\p}{\partial}

\newcommand{\bea}{\begin{equation}\begin{aligned}}
\newcommand{\eea}{\end{aligned}\end{equation}}
\newcommand{\ba}{\begin{align}}
\newcommand{\ea}{\end{align}}

\newcommand{\ra}{\rangle}
\newcommand{\la}{\langle}
\begin{document}

\begin{titlepage}

\vspace{.4cm}
\begin{center}
\noindent{\Large \textbf{Relative entropy of excited states in conformal field theories of arbitrary dimensions}}\\
\vspace{1cm}
G\'abor S\'arosi$^{a,b,}$\footnote{gsarosi@vub.ac.be}
and
 Tomonori Ugajin$^{c,}$\footnote{ugajin@kitp.ucsb.edu}

\vspace{.5cm}
 {\it
 $^{a}$Theoretische Natuurkunde, Vrije Universiteit Brussels and \\ International Solvay Institutes,\\
Pleinlaan 2, Brussels, B-1050, Belgium\\
\vspace{0.2cm}
 }
 \vspace{.5cm}
  {\it
  $^{b}$David Rittenhouse Laboratory, University of Pennsylvania,\\
  Philadelphia, PA 19104, USA\\
\vspace{0.2cm}
 }
\vspace{.5cm}
  {\it
 $^{c}$Kavli Institute for Theoretical Physics, University of California, \\
Santa Barbara, 
CA 93106, USA\\
\vspace{0.2cm}
 }
\end{center}


\begin{abstract}
Extending our previous work, we study the relative entropy between the reduced density matrices obtained from globally excited states in conformal field theories of arbitrary dimensions. We find a general formula in the small subsystem size limit.  When one of the states is the vacuum of the CFT, our result matches with the holographic entanglement entropy computations in the corresponding bulk geometries, including AdS black branes. We also discuss the first asymmetric part of the relative entropy and comment on some implications of the results on the distinguishability of black hole microstates in AdS/CFT.
\end{abstract}

\end{titlepage}

\tableofcontents
\pagebreak

\section{Introduction}

The relative entropy $S(\rho || \sigma)$ is a useful statistical measure to distinguish two density matrices $ \rho$ and $\sigma$.  It is defined as 
\be
S( \rho || \sigma) = {\rm tr} \rho \log \rho  - {\rm tr} \rho \log \sigma.
\ee
Among its most important properties are that it is always positive (zero iff $\sigma=\rho$) and that it decreases under completely positive trace preserving maps (monotonicity).
The operational meaning of the relative entropy is the following. The probability of the results of $n$ measurements performed in the state $\rho$ being accurately described by a different density matrix $\sigma$ is $e^{-nS(\rho ||\sigma)}$. Notice that the relative entropy is not symmetric in its arguments\footnote{A very nice example to understand this is given in \cite{Vedral:2002zz}: if we are given a fair coin, after a couple of tosses we can be sure that the coin is not giving heads with probability one. This is not true the other way around, if we are given the coin which gives heads all the times, we can never be entirely sure that it is unfair.} and therefore it is not a metric distance between the states. This reflects the fact that states that are close in some metric distance can still be easily distinguishable by measurements. Indeed, the relative entropy can be used to bound from above several metric distances, such as the trace distance \cite{ohya2004quantum}. We refer to \cite{Vedral:2002zz} for an extensive review of the relative entropy.

\vspace{0.2cm}

In recent years, the relative entropy paved its way to become a successful tool in quantum field theory and quantum gravity. Its properties were used to understand the Bekenstein bound \cite{Casini:2008cr} and to prove the quantum Bousso bound \cite{Bousso:2014sda,Bousso:2014uxa}, the generalized second law \cite{Wall:2011hj} and the average null energy condition in Minkowski space \cite{Faulkner:2016mzt}. In the holographic set up, positivity of the relative entropy can be used for example to derive the linearized bulk equations of motion \cite{Faulkner:2013ica} or to obtain constraints on gravitational systems with a unitary holographic description \cite{Lin:2014hva,Lashkari:2014kda,Lashkari:2016idm}. Despite all this success, there are relatively few direct calculations of the relative entropy available in a purely quantum field theoretic set up \cite{Lashkari:2014yva,Lashkari:2015dia}.

\vspace{0.2cm}

In order to make progress in this latter direction, in our previous work \cite{Sarosi:2016oks}, we have studied the relative entropy between the reduced density matrices $\rho_{V}$ and $\rho_{W}$ of primary energy eigenstates $|V \ra$, $|W \ra$ 
\bea
\rho_{V} &={\rm tr}_{A^{c}} |V \ra \la V| , && \rho_{W} &={\rm tr}_{A^{c}} |W \ra \la W|, \label{eq: redu}
\eea
associated with a single interval $A$ in the case of two dimensional conformal field theories.  Here, $A^{c}$ denotes the complement of the subsystem $A$. We have obtained the following general formula for the relative entropy, to leading order in the size of the subsystem $|A| = 2\pi x \rightarrow 0$
\be
S(\rho_{V}||\rho_{W})=\frac{\Gamma(\f{3}{2})\Gamma(\Delta+1)}{2\Gamma(\Delta+\f{3}{2})}\sum_{\alpha}\left(C_{O_{\alpha}VV}-C_{O_{\alpha}WW} \right)^2 (\pi x)^{2\Delta}+\cdots,  \label{eq:relent1}
\ee
where $\{O_{\alpha} \}$ is the set of lightest operators\footnote{These do not have to be primaries, one of them can be the stress tensor, though we need to keep in mind that the sum over $O_\alpha$ must be understood as a contraction with the Zamolodchikov metric $g^{O_\alpha O_\alpha}$. } with $C_{VVO_{\alpha}}-C_{WWO_{\alpha}} \neq 0$ and $\Delta$ is the scaling dimension of these operators.
To derive this result, we have used the replica trick proposed in \cite{Lashkari:2015dia}.  We have checked this result against some
known special cases \cite{Lashkari:2015dia, Blanco:2013joa} and to a computation involving generalized free fields. One puzzling aspect of this latter computation is that the formula appears to be independent of the spacetime dimensionality. This suggests that \eqref{eq:relent1} could hold as it stands in an arbitrary number of dimensions.

\vspace{0.2cm}

In this paper we compute the leading order relative entropy between two excited states reduced to a single ball shaped region in higher dimensional conformal field theories\footnote{Our assumption about the theories is that they can be coupled to a background metric in a diffeomorphism invariant way and that the resulting theory is (at least classically) invariant under Weyl transformations $g\rightarrow \Omega^2g$ of the background metric. This is a stronger assumption than global conformal invariance in flat space whenever $d>2$ \cite{Nakayama:2013is,Karananas:2015ioa}. We will nevertheless refer to these theories conformal field theories in the following.} defined on $\mathbb{R} \times S^{d-1}$. We employ again the replica trick in \cite{Lashkari:2015dia} to do this. 
By using the Casini-Huerta-Myers (CHM) map \cite{Casini:2011kv}, the computation boils down to the calculations of certain correlation functions  on $ S^{1} \times H^{d-1}$, where $H^{d-1}$ is the $d-1$ dimensional hyperbolic space. 
When the dominant operator $O$ is a scalar we find that the leading order relative entropy agrees with formula \eqref{eq:relent1} for 2d CFTs with $x$ being replaced by $\frac{\theta_0}{\pi}$, with $\theta_0$ being the radius of the entangling sphere. Note the remarkable feature that this result is indeed independent of the spacetime dimensionality. Contrary to this, when the dominant operator is the stress tensor, then the leading order result is
\be 
\label{eq:higherdimstressrel}
S( \rho_{V} || \rho_{W}) =\f{1}{4C_{T}} \left(\f{d}{d-1} \right) ( \varepsilon_{V}-\varepsilon_{W})^2 \f{\Gamma(\f{1}{2}) \Gamma(d+1)}{\Gamma(d+\f{3}{2})}\; \theta_{0}^{2d} + \cdots,
\ee
where $C_{T}$ is the coefficient of the stress tensor two point function, and $\varepsilon$ denotes the energy density, i.e. the scaling dimension divided by  the volume of the spatial sphere, $S^{d-1}$. We will also give the formula in the case when the dominant operator is a conserved $U(1)$ current.

\vspace{0.2cm}
When one of the states is the vacuum of the CFT we can compare our formulas to the holographic results of \cite{Blanco:2013joa}. There, the holographic entanglement entropy \cite{Ryu:2006bv,Ryu:2006ef} is calculated on the AdS black brane background and on a background with a bulk scalar condensate  
up to second order in the deformations compared to pure AdS in the small subsystem size limit. This is used to infer the relative entropy between the vacuum and an excited state. We find that our CFT results perfectly agree with these holographic results. It is worth to note that to derive the holographic entanglement entropy at second order, one needs the nonlinear parts of Einstein's equations. This is to be contrasted with the first order result which is needed to confirm the first law of entanglement and only uses the linearized field equations. This suggests that the general CFT 
results know about the nonlinear parts of the bulk Einstein's equations. 

\vspace{0.2cm}

As we have already mentioned, the relative entropy is in general not symmetric under the exchange of the two density matrices in its arguments. However, the leading order results (\ref{eq:relent1}), (\ref{eq:higherdimstressrel}) are symmetric. We may interpret this as a consequence of all energy eigenstates resembling the vacuum on short distances and therefore we are essentially calculating the norm of the difference of the two reduced density matrices in the Fisher information metric (see e.g. \cite{nielsen2010quantum}) corresponding to the vacuum. The first asymmetric part of the relative entropy appears at next to leading order in the expansion in the subsystem size. We will also discuss the dependence of this contribution on the OPE coefficients. 
In particular, we consider a class of excited states in 2d CFTs which behave as thermal states in the thermodynamic limit, i.e. when the size of the subsystem is much smaller than the length of the thermal cycle. These have a local expression for their modular Hamiltonian and we confirm that the expansion of the resulting relative entropy reproduces the asymmetric part that we have found by general considerations.

\vspace{0.2 cm}

While the leading result \eqref{eq:relent1} is in general not universal, there are some statements that can be made about its magnitude. This, in particular, gives us information about the distinguishability of black hole microstates in the context of AdS/CFT \cite{Maldacena:1997re,Witten:1998qj}. Indeed, the results \eqref{eq:relent1}, \eqref{eq:higherdimstressrel} can be interpreted as the bulk relative entropy in the entanglement wedge of region $A$ in holographic theories \cite{Jafferis:2015del}. As an example we will show, provided that the lightest operator is the stress tensor, that in large $N$ theories heavy states within an energy window of $\frac{\Delta E}{E} \lesssim \frac{1}{N}$ are indistinguishable inside the region $A$ to leading order in $\frac{1}{N}$.
Another example is given by theories satisfying local eigenstate thermalization -- which is defined as the thermalization of all the local one point functions in a high energy eigenstate\footnote{A recent calculation \cite{Lin:2016dxa} shows that the R\'enyi entanglement entropies are different for the thermal state and highly excited states even in the small interval expansion in 2d CFTs. This somewhat undermines the assumption of local ETH but it appears to us that the differences discussed in \cite{Lin:2016dxa} go away in the large $c$ limit.}. It was argued very recently in \cite{Lashkari:2016vgj} that under this assumption the heavy-heavy-light OPE coefficients are universal up to corrections suppressed by $e^{-O(S(E))}$, where $S(E)$ is the microcanonical entropy at the heavy state energy $E$. This was used to show that the relative entropy between the reduced density matrices of excited states and a universal density matrix is small, proportional to $e^{-O(S(E))}$. Here we note that the same assumption on OPE coefficients implies via e.q. \eqref{eq:relent1} that the relative entropy between reduced density matrices of different primary states with the same energy is also of order $e^{-O(S(E))}$. Note that while these two statements are intuitively equivalent, they do not follow from each other as the relative entropy does not satisfy any triangle inequality.

\vspace{0.2 cm}

The organization of this paper is as follows. 
In section \ref{section:setup} we explain our setup and how to calculate the relative entropy in terms of correlation functions on $ S^{1} \times H^{d-1}$. In section \ref{section:mainresults} we derive our main results by expanding the correlation functions in the small subsystem limit $\theta_{0} \rightarrow 0$. We also explain here how to understand our formulas using the point-like limit of the excited state modular Hamiltonian. 
In section \ref{section:holography} we compare our CFT results with the holographic results of \cite{Blanco:2013joa}. We find a perfect agreement. In section \ref{sec:beyondsmall} we discuss the leading asymmetric contribution to the relative entropy and compare it to a two dimensional example. In section \ref{sec:variance} we examine the average relative entropy between states within a given energy window. In appendix \ref{app:ope} and  \ref{app:analcont} we 
explain some details of our calculations.

\section{Preliminaries}
\label{section:setup}

In this section we explain how to compute the relative entropy $S( \rho_{V} || \rho_{W})$ between two globally excited energy eigenstates  $| V \ra$, $|W \ra$ reduced to a single ball shaped region in conformal field theories of generic dimension.  We do this by employing the replica trick \cite{Lashkari:2015dia}
\begin{align}
 S(\rho_{V}||\rho_{W})&=\lim_{n \rightarrow 1} S_{n}(\rho_{V}||\rho_{W}) \nonumber \\ 
&=\lim_{n \rightarrow 1} \f{1}{n-1} \left( \log {\rm tr} \;\rho_{V}^n- \log{\rm tr} \;\rho_{V} \rho_{W}^{n-1} \right) \label{eq:relrep}
\end{align}
and writing down the two terms in the replica relative entropy (\ref{eq:relrep}) in terms of correlation functions on $S^{1} \times H^{d-1}$.

Let us first look at  the R\'enyi entropies $ {\rm tr}\; \rho_{V}^n$ of an excited state $| V \ra$.
We start from a CFT on $\mathbb{R} \times S^{d-1}$, on which we use the following metric
\be
ds_{\mathbb{R} \times S^{d-1}}^2= dt^2 +(d \theta^2 +\sin ^2 \theta\; d \Omega^{2}_{d-2}). \label{eq:metric}
\ee
 We define our subsystem $A$ to be a cap like region $[ 0, \theta_{0}]$. Then, $ {\rm tr} \rho_{V}^n$ is computed by the path integral on the $n$ sheet copy of this manifold with a cut on the subsystem $A$ and the excited state $|V \rangle$ located at $t= \pm \infty$ of each sheet. Now we briefly describe the conformal maps which bring this branched cylinder to the covering manifold $S^1_n\times H^{d-1}$, where the circle has periodicity $2\pi n$.

\begin{description}
\item[From branched cylinder to branched sphere $S^{d}_{n}$:]

Each cylinder sheet $\mathbb{R} \times S^{d-1}$ can be mapped to a $d$-dimensional sphere $S^{d}$ with the coordinates $( \phi , \Omega_{d-1})$ by the conformal map

\be
t= \log \tan \f{\phi}{2}.
\ee
Then,

\be
ds_{\mathbb{R} \times S^{d-1}}^2=\f{1}{\sin ^2\phi} ds_{S^{d}}^2 \equiv \Omega_{1}^2\; ds_{S^{d}}^2, \qquad ds_{S^{d}}^2 =d\phi ^2 +\sin^2 \phi \; d \Omega_{d-1}^2.
\ee 

The excited states on the $k$-th cylinder sheet are mapped to the north pole $ \phi= \pi_{k} $ and  the south pole $ \phi= 0_{k}$ of the $k$-th sheet of the sphere. By the state operator correspondence, the excited states on these poles can be written in terms of a local operator

\be
\langle V| = \langle 0|  V(\pi)   \qquad |V \rangle = V(0) |0 \rangle.
\ee
By using this, we can write the R\'enyi entropy as
\be
\label{eq:Renyiratio}
 {\rm tr} \;\rho_{V}^n =\mathcal{N}_n\f{\langle \prod_{k=0}^{n-1} V(\pi_{k}) V(0_{k}) \rangle_{S^{d}_{n}}}{ \prod_{k=0}^{n-1}  \langle V(\pi_{k}) V(0_{k}) \rangle_{S^{d}}},
\ee
where $\mathcal{N}_n$ is some regularization dependent normalization constant involving the partition function on $S^d_n$ and $S^d$ but independent of the operator $V$. We divide by the appropriate power of $\langle V|V\rangle=\langle V(\pi)V(0)\rangle_{S^d}$ to have normalized states.
Note that the $V$ dependent part on the right hand side of \eqref{eq:Renyiratio} is invariant under the Weyl transformation $g_{\mu \nu} \rightarrow \Omega^2 g_{\mu \nu}$.

\item[From  branched sphere  $S^{d}_{n}$ to $\Sigma_{n} \equiv S^{1}_{n} \times H^{d-1}$:]

Now we introduce uniformized coordinates by conformally mapping the branched sphere $S^{d}_{n}$ to the manifold $\Sigma_{n} \equiv S^{1}_{n} \times H^{d-1}$, where $H^{d-1}$ is the hyperbolic space. Notice again that we do not need to keep track of the Weyl factors as these cancel out from the expression for $ {\rm tr} \;\rho_{V}^n$.  The coordinates on $\Sigma_n$ are $( \tau, u,  { \Omega_{d-2} })$ with the identification $\tau \sim \tau +2n \pi$ and the metric is
\be
ds^2_{\Sigma_{n}}= d\tau^2 + du^2+{ \sinh^2 u d\Omega_{d-2}^2}.
\ee
The precise expression of the map is given by \cite{Casini:2011kv}
\be
-\cos \phi =\f{\sin \theta_{0} \sin \tau}{\cosh u+ { \cos \theta_0}\cos \tau}, \qquad \tan \theta =\f{\sin \theta_{0}\sinh u}{\cos \tau+\cos \theta_{0} { \cosh u} }.
\ee
The local operators on $S^{d}_{n}$ are mapped to the following positions of   $\Sigma_{n}$  
\bea
\label{eq: opeloc}
\pi_{k} \rightarrow& \tau_{k}=2\pi k +(\pi-\theta_{0}), \quad u=0,  \\
0_{k} \rightarrow& \hat{\tau_{k}}=2\pi k +(\pi+\theta_{0}), \quad u=0. 
\eea
Note that in the small interval limit, $ \theta_{0} \rightarrow 0$, $\tau_{k} \rightarrow \hat{\tau_{k}}$, i.e. the insertions pairwise approach each other.

\end{description}

The trace of the reduced density matrix  ${\rm tr} \;\rho_{V}^n$ can again be expressed by insertions of a local operator $V(\tau,u)$ on the manifold 
\be
{\rm tr} \;\rho_{V}^n=\mathcal{N}_n\f{\langle \prod_{k=0}^{n-1} V(\tau_{k},0) V(\hat{\tau}_{k},0) \rangle_{\Sigma_{n}}}{\prod_{k=0}^{n-1}  \langle V(\tau_{k}) V(\hat{\tau}_{k}) \rangle_{\Sigma_{1}}} \label{eq:rep3},
\ee 
Similarly, the second term in (\ref{eq:relrep}) is computed by the correlation function
\be
{\rm tr} \;\rho_{V} \rho_{W}^{n-1}=\mathcal{N}_n\f{\langle V(\tau_{0},0) V(\hat{\tau}_{0},0)\prod_{k=1}^{n-1} W(\tau_{k},0) W(\hat{\tau}_{k},0) \rangle_{\Sigma_{n}}}{  \langle V(\tau_{0}) V(\hat{\tau}_{0}) \rangle_{\Sigma_{1}}\prod_{k=1}^{n-1}  \langle W(\tau_{k}) W(\hat{\tau}_{k}) \rangle_{\Sigma_{1}}} \label{eq:rep2}.
\ee
These two quantities involve the same normalization constant $ \mathcal{N}_n$ and we see that this cancels from the \eqref{eq:relrep} definition of the replica relative entropy. Therefore, in the following we set $\mathcal{N}_n=1$.

\section{Relative entropy in the small radius limit}
\label{section:mainresults}

\subsection{Scalar exchange}
\label{subsection:firstlawtrick}
\noindent
In this subsection we derive the leading term in the small radius limit $\theta_{0} \rightarrow 0$, when the lightest operator $O$ in the $ V \times V$ OPE on $\Sigma_{n}$  
\be
V( \tau_{k}) V (\hat{\tau}_{k})=
\la V( \tau_{k}) V (\hat{\tau}_{k})\ra_{\Sigma_{n}} \Bigl\{1+ C_{VV}^{O} D_{n}( \tau_{k}, \hat{\tau_{k}})^{\Delta_{O}} O(\tau_{k})  + \cdots \Bigr\}.
\label{eq:scalarope}
\ee
is a scalar. In the following, we only highlight the $\tau$ arguments of operators as the insertion positions on $H^{d-1}$ are the same.
Note that we do not know the explicit form of either $\la V( \tau_{k}) V (\hat{\tau}_{k})\ra_{\Sigma_{n}}$ or $D_{n}( \tau_{k}, \hat{\tau_{k}})$ when $n \neq 1$. However, as we will see below, to calculate the relative entropy we only need their expressions for $n=1$. We also note that because of translational invariance along the $\tau$ direction, these two functions only depend on the differences of $\tau$ ie, $\la V( \tau_{k}) V (\hat{\tau}_{k})\ra_{\Sigma_{n}} = G_{n}(\tau_{k}-\hat{\tau}_{k})$ and $D_{n}( \tau_{k}, \hat{\tau_{k}}) =D_{n}(\tau_{k}-\hat{\tau}_{k})$. In particular, they are independent of $k$.
Using the OPE \eqref{eq:scalarope} in \eqref{eq:rep3} we get at leading order
\be
\label{eq:trrhovn}
{\rm tr} \;\rho_{V}^n=\left(\f{ \langle V(\tau_{k}) V(\hat{\tau}_{k}) \rangle_{\Sigma_{n}}}{ \langle V(\tau_{k}) V(\hat{\tau}_{k}) \rangle_{\Sigma_{1}}} \right)^{n}\left(1+(C_{VV}^O)^2 D_{n}( \tau - \hat{\tau})^{2\Delta_O}   \frac{1}{2}\sum_{k\neq l=0}^{n-1} \langle O( \tau_{k})  O(\tau_{l})  \rangle_{\Sigma_{n}} + \cdots  \right)
\ee
Similarly 
\begin{align}
\label{eq:trrhovwn}
{\rm tr} \;\rho_{V} \rho_{W}^{n-1}&=\f{ \langle V(\tau_{0}) V(\hat{\tau}_{0}) \rangle_{\Sigma_{n}}  \left(\langle W(\tau_{k}) W(\hat{\tau}_{k}) \rangle_{\Sigma_{n}} \right)^{n-1}}{\langle V(\tau_{0}) V(\hat{\tau}_{0}) \rangle_{\Sigma_{1}}\left(  \langle W(\tau_{k}) W(\hat{\tau}_{k}) \rangle_{\Sigma_{1}}\right)^{n-1}} \nonumber \\[+10pt]
&\times \left( 
1+C_{VV}^{O} C_{WW}^{O}  D_{n}( \tau, \hat{\tau})^{2\Delta_O} \sum_{ l=1}^{n-1}  \langle O( \tau_{0})  O(\tau_{l})  \rangle_{\Sigma_{n}} \right. \nonumber \\
 &\left.+(C_{WW}^{O})^2 D_{n}( \tau, \hat{\tau})^{2\Delta_O}  \frac{1}{2}\sum_{k\neq l =1}^{n-1}  \langle O( \tau_{k}) O(\tau_{l})  \rangle_{\Sigma_{n}}+\cdots\right).
\end{align} 
We can simplify the double sums in these expressions as
\bea
\sum_{k\neq l=0}^{n-1} \langle O( \tau_{k})  O(\tau_{l})  \rangle_{\Sigma_{n}} &= \sum_{k\neq l=0}^{n-1} \langle O( \tau_{k}-\tau_{l})  O(0)  \rangle_{\Sigma_{n}}  \\
&=\sum_{k\neq l=0}^{n-1} \langle O(2\pi(k-l))  O(0)  \rangle_{\Sigma_{n}} \\
&=\sum_{k=0}^{n-1}\sum_{m=k-n+1,m\neq 0}^{k}\langle O(2\pi m)  O(0)  \rangle_{\Sigma_{n}} \\
&= \left( \sum_{m=1-n}^{-1}\sum_{k=0}^{m+n-1} + \sum_{m=1}^{n-1}\sum_{k=m}^{n-1}\right)\langle O(2\pi m)  O(0)  \rangle_{\Sigma_{n}} \\
&=2\sum_{m=1}^{n-1}(n-m)\langle O(2\pi m)  O(0)  \rangle_{\Sigma_{n}} ,
\eea
where we are using translational invariance in the $\tau$ direction in the first line, the expression \eqref{eq: opeloc} for $\tau_k$ in the second line, change of the summation variable to $m=k-l$ in the third line, exchange of sums in the fourth line and invariance of the correlator under $\tau \rightarrow -\tau$ in the fifth line. Finally, we introduce $k=m-n$ as a new summation variable and use that the $\tau$ direction is $2\pi n$ periodic to arrive at
\beq
\sum_{m=1}^{n-1}(n-m)\langle O(2\pi m)  O(0)  \rangle_{\Sigma_{n}} =\sum_{k=1}^{n-1}k\langle O(2\pi k)  O(0)  \rangle_{\Sigma_{n}},
\eeq
from which, after rearrangement, it also follows that
\beq
\sum_{k=1}^{n-1}k\langle O(2\pi k)  O(0)  \rangle_{\Sigma_{n}}=\frac{n}{2}\sum_{k=1}^{n-1}\langle O(2\pi k)  O(0)  \rangle_{\Sigma_{n}}.
\eeq
Combining \eqref{eq:trrhovn} and \eqref{eq:trrhovwn} and using the above simplification for the double sums we get
\begin{align}
S_{n}(\rho_{V}|| \rho_{W})& =\frac{1}{n-1}\left(\log \f{ \langle V(\tau_{k}) V(\hat{\tau}_{k}) \rangle_{\Sigma_{n}}}{ \langle V(\tau_{k}) V(\hat{\tau}_{k}) \rangle_{\Sigma_{1}}}-\log \f{ \langle W(\tau_{k}) W(\hat{\tau}_{k}) \rangle_{\Sigma_{n}}}{ \langle W(\tau_{k}) W(\hat{\tau}_{k}) \rangle_{\Sigma_{1}}}\right)  \nonumber \\[+10pt]
&=Q_n+\f{f(\Delta_O,n)}{n-1}\left( \f{n}{2} (C^{O}_{VV})^2 -C^{O}_{WW}\;C^{O}_{VV}-\f{(n-2)}{2}(C^{O}_{WW})^2 \right) D_{n}( \tau, \hat{\tau})^{2\Delta_O}+\cdots,
\end{align}
where $f(\Delta_O,n)$ is given by 
\be
f(\Delta_O,n)= \sum_{k=1}^{n-1} \la O( 2\pi k) O (0)\ra_{\Sigma_{n}},
\ee
and $Q_n=\log \left( \frac{\langle W(\tau_{0}) W(\hat{\tau}_{0}) \rangle_{\Sigma_{1}}}{\langle V(\tau_{0}) V(\hat{\tau}_{0}) \rangle_{\Sigma_{1}}}\frac{\langle V(\tau_{0}) V(\hat{\tau}_{0}) \rangle_{\Sigma_{n}}}{\langle W(\tau_{0}) W(\hat{\tau}_{0}) \rangle_{\Sigma_{n}}}\right)$. It is clear that $\lim_{n\rightarrow 1} Q_n=0$.
The analytic continuation of $f(\Delta,n)$ can be done using the methods developed in \cite{Agon:2015ftl}. We present some details of this in appendix \ref{app:analcont} and quote the final result
\be
\lim_{n \rightarrow 1} \f{f(\Delta,n)}{n-1} =\f{\Gamma(\f{1}{2})\; \Gamma(\Delta+1)}{2^{2\Delta+1}\Gamma( \Delta+\f{3}{2})}.
\ee 
The last piece we need is the expression for $D_{n=1}( \tau-\hat{\tau})$. Although it is possible to determine this explicitly, here we only need the leading behavior when $\tau \rightarrow \hat \tau$, i.e. when the insertions approach each other. In this case every smooth manifold should look like flat space, therefore we may write
\bea
D_{n=1}( \tau-\hat{\tau})& \approx |\tau-\hat{\tau}| + O((\tau-\hat{\tau})^2)\\
&=2\theta_0 + O(\theta_0^2).
\eea
Putting the pieces together we obtain for the relative entropy the expression
\be
\label{eq:relscalar}
S(\rho_{V}|| \rho_{W})=\f{\Gamma(\f{1}{2}) \Gamma(\Delta_O+1)}{4\Gamma( \Delta_O+\f{3}{2})} (C^{O}_{VV}- C^{O}_{WW})^{2} \theta_{0}^{2\Delta_O } + \cdots.
\ee
An interesting feature of this formula is that it is independent of the dimension $d$. This will not be the case when the leading exchange is not a scalar operator.

\subsubsection{Modular Hamiltonian}

A different way to understand the result \eqref{eq:relscalar} is the following.
By doing the analytic continuation of the R\'enyi entropy \eqref{eq:trrhovn} alone we see that the entanglement entropy of the excited state $|W \rangle $ in the small interval limit is given by\footnote{See \cite{Alcaraz:2011tn} for the two dimensional case.}
\be
S_{A}(\rho_{W}) =S_{\rm vac} + S_{\text{first law}} -\f{\Gamma(\f{1}{2}) \Gamma( \Delta_O+1)}{4 \Gamma(\Delta_O+\f{3}{2})}\; \theta_0^{2\Delta_O} \; \langle W|O|W \rangle^2.
\ee
Here, $S_{\rm vac}$ is the entanglement entropy in the vacuum state (containing all the UV divergences) and 
\beq
S_{\text{first law}}=\lim_{n\rightarrow 1}\frac{n}{1-n} \log \left(\f{ \langle W(\tau_{k}) W(\hat{\tau}_{k}) \rangle_{\Sigma_{n}}}{ \langle W(\tau_{k}) W(\hat{\tau}_{k}) \rangle_{\Sigma_{1}}} \right)
\eeq
is the completely factorized contribution which drops out from the relative entropy.
Now let us deform  the reduced density matrix slightly $ \rho_{W} \rightarrow \rho_{W}+ \delta \rho$. Then the entanglement entropy $ S_{A}$ changes as
\be
\delta S_{A} =\delta S_{\text{first law}}-2c_{A} \langle  W|O|W\rangle \text{Tr}( O \delta \rho ) , \quad c_{A} \equiv \f{\Gamma(\f{1}{2}) \Gamma( \Delta_O+1)}{4\Gamma(\Delta_O+\f{3}{2})}\; \theta_0^{2\Delta_O} .
\ee
At the same time, because of the first law of entanglement, we have
\be 
\label{eq:variationaltrick}
\delta S_{A}= \text{Tr}( K^{W}_{A} \delta \rho ), \quad K^{W}_{A} = -\log \rho_{W}.
\ee 
We claim that $\delta S_{\text{first law}}=\text{Tr}(K^{\rm vac}_A \delta \rho)$, where $K^{\rm vac}_A$ is the modular Hamiltonian associated to the vacuum. Indeed, using \eqref{eq:rep2} with one of the operators being the identity we have
\beq
\mathcal{N}_n \f{ \langle W(\tau_{k}) W(\hat{\tau}_{k}) \rangle_{\Sigma_{n}}}{ \langle W(\tau_{k}) W(\hat{\tau}_{k}) \rangle_{\Sigma_{1}}}  = \text{Tr} \rho_W \rho_{\rm vac}^{n-1},
\eeq
therefore
\beq
S_{\rm vac}+S_{\text{first law}}=-\text{Tr}\rho_W\log\rho_{\rm vac} \equiv \text{Tr}\rho_W K^{\rm vac}_A.
\eeq
Since equation \eqref{eq:variationaltrick} holds for any deformation of the reduced density matrix we can read off the expression for the modular Hamiltonian\footnote{We thank Stefan Leichenauer for discussion on this. See also \cite{Jafferis:2015del,Lashkari:2016vgj} for recent applications.} $K^{W}_{A}$ 
\be
K^{W}_{A} =K^{\rm vac}_A-2c_{A} \langle W| O|W \rangle O \label{eq:modex}
\ee 
valid to leading order in the small subsystem size limit $\theta_0 \rightarrow 0$. In the above expression, the local operator $O$ is inserted at an arbitrary point inside $A$ to leading order in the size of $A$. Note that by dimensional analysis we have $K^{\rm vac}_A \sim \theta_0^d T$ in the small interval limit, where $T$ is the stress tensor. Therefore the modular Hamiltonian is dominated by the second term in the short interval limit only when $\Delta_O < \frac{d}{2}$, otherwise it is dominated by the vacuum modular Hamiltonian.
Regardless of this, we can evaluate the relative entropy using expression \eqref{eq:modex} as
\begin{align}
S(\rho_{V}|| \rho_{W}) &= \Delta \langle K^{W}_{A} \rangle - \Delta S \nonumber \\
&=c_{A} \Bigl( \langle V|O|V \rangle-\langle W |O| W\rangle \Bigr)^2,
\end{align}
which, of course, agrees with \eqref{eq:relscalar}.

\subsection{Stress tensor exchange} 
In this subsection, we consider the contribution of stress tensor exchange to the relative entropy in the small radius $\theta_0 \rightarrow 0$ limit. In the OPE of two operators $V(\tau_{k}), V(\hat{\tau}_{k})$ on $\Sigma_{n}$, the stress tensor $T_{MN}$ appears as
\be
\label{eq:OPEstress}
V (\tau_{k})V (\hat{\tau}_{k}) =\langle V (\tau_{k})V (\hat{\tau}_{k}) \rangle_{\Sigma_{n}} \left[1+ {C}^{MN}_{VV} (\Sigma_n:\tau-\hat{\tau}) T_{MN} (\tau_{k}) + \cdots \right],
\ee
where the details of the functions ${C}^{MN}_{VV} (\Sigma_n:\tau-\hat{\tau})$ depend on the manifold $\Sigma_n$. An important caveat to this expansion is that not all of the identity operator contributions are taken into account by the first term. Indeed, when there is a nonzero stress tensor expectation value $\langle T_{MN}(\tau_k)\rangle_{\Sigma_n}$ due to conformal anomaly, we must understand the second term as
\be
V (\tau_{k})V (\hat{\tau}_{k}) =\langle V (\tau_{k})V (\hat{\tau}_{k}) \rangle_{\Sigma_{n}} \left[1+ {C}^{MN}_{VV} (\Sigma_n:\tau-\hat{\tau}) \left( T_{MN} (\tau_{k})-\langle T_{MN} (\tau_{k})\rangle_{\Sigma_n}\right) + \cdots \right],
\ee
so that the expectation value of the left and right hand sides coincide. In the following we will abuse notation and just write $T_{MN} $ keeping in mind that we do not need to worry about anomalous terms in the correlation functions involving $T_{MN}$.
  
  Inserting \eqref{eq:OPEstress} into \eqref{eq:rep2} and \eqref{eq:rep3} and repeating the steps from the previous section we get the following leading order replica relative entropy
\begin{align}
S_{n}(\rho_{V}|| \rho_{W})& =\frac{1}{n-1}\left(\log \f{ \langle V(\tau_{k}) V(\hat{\tau}_{k}) \rangle_{\Sigma_{n}}}{ \langle V(\tau_{k}) V(\hat{\tau}_{k}) \rangle_{\Sigma_{1}}}-\log \f{ \langle W(\tau_{k}) W(\hat{\tau}_{k}) \rangle_{\Sigma_{n}}}{ \langle W(\tau_{k}) W(\hat{\tau}_{k}) \rangle_{\Sigma_{1}}}\right)  \nonumber \\[+10pt]
&=Q_n+\f{\sigma_{n}(AB,MN)}{n-1}\left( \f{n}{2} C^{AB}_{VV}  C^{MN}_{VV}   -C^{AB}_{WW} \;C^{MN}_{VV}  -\f{(n-2)}{2} C^{AB}_{WW} C^{MN}_{WW} \right) +\cdots,
\end{align} 
where
\beq
\sigma_{n} (AB,MN) = \sum_{k=1}^{n-1} \langle T_{MN}( \tau_{k}) T_{AB}(\tau_{0}) \rangle_{\Sigma_n},
\eeq
and summation over the double indices $A,B,M,N$ is understood.
Again, since $\sigma_{n}(AB,MN)$ is proportional to $n-1$ we only need to evaluate $C^{MN}_{VV}$ on $\Sigma_{1}$. We present the details of the calculations of $C^{MN}_{VV}$ and $\sigma_{n}(AB,MN)$ in appendix \ref{app:stressope}, and \ref{app:analcont} respectively. The results are
\bea
\lim_{n\rightarrow 1}\frac{1}{n-1}\sigma_{n} (\tau \tau, \tau \tau) &=  \f{C_{T}}{2^{2d+1}} \f{d-1}{d} \f{\Gamma (\f{1}{2}) \Gamma(d+1)}{\Gamma(d+\f{3}{2})},\\
C^{MN} (\Sigma_1 : \tau-\hat \tau)&=\f{d \varepsilon_V}{1-d}\f{1}{ C_{T}}\left( 2\sin \theta_{0} \right)^{d} \delta^{\tau M}  \delta^{\tau N} ,
\eea
where $C_{T}$ is the coefficient of the two point function of the stress tensor and $\varepsilon_{V}$ is the energy density of the excited state $|V \ra$
\be
\varepsilon_{V}=\f{\Delta_{V}}{\Omega_{d-1}}, \quad \Omega_{d-1}= \f{2 \pi^{\f{d}{2}}}{\Gamma \left( \f{d}{2} \right)}.
\label{eq:dsphere}
\ee

\noindent
By putting everything together we obtain the main result of this subsection
\be
S(\rho_{V}|| \rho_{W}) =\f{1}{4C_{T}} \left(\f{d}{d-1} \right) (\varepsilon_{V}-\varepsilon_{W})^2 \f{\Gamma (\f{1}{2}) \Gamma(d+1)}{\Gamma(d+\f{3}{2})} (\sin \theta_{0})^{2d} +\cdots \label{eq:cftrel}.
\ee
For the case $d=2$ we can use $\theta_0=\pi x$, $C_T=\frac{c}{2\pi^2}$ (here $c$ is the Virasoro central charge)\footnote{We can match normalization of the stress tensor by writing the two point funtion of the Euclidean time components $T_{00}=-\frac{1}{2\pi}(T(z)+\bar T(\bar z))$, where $T$ and $\bar T$ are the usual holomorphic and antiholomorphic stress tensors in 2d \cite{DiFrancesco:1997nk}.} and $\varepsilon_V=\frac{h_V}{\pi}$ to recover the result of \cite{Sarosi:2016oks}.

\subsection{$U(1)$ current exchange}

For completeness, we include the case when the dominant contribution in the $V\times V$ OPE is a $U(1)$ current $J$, i.e.
\be
\label{eq:OPEstress}
V (\tau_{k})V (\hat{\tau}_{k}) =\langle V (\tau_{k})V (\hat{\tau}_{k}) \rangle_{\Sigma_{n}} \left[1+ {C}^{M}_{VV} (\Sigma_n:\tau-\hat{\tau}) J_{M} (\tau_{k}) + \cdots \right],
\ee
The calculation proceeds in an entirely analogous manner as in the previous two cases. We have for the replica relative entropy
\begin{align}
S_{n}(\rho_{V}|| \rho_{W})& =\frac{1}{n-1}\left(\log \f{ \langle V(\tau_{k}) V(\hat{\tau}_{k}) \rangle_{\Sigma_{n}}}{ \langle V(\tau_{k}) V(\hat{\tau}_{k}) \rangle_{\Sigma_{1}}}-\log \f{ \langle W(\tau_{k}) W(\hat{\tau}_{k}) \rangle_{\Sigma_{n}}}{ \langle W(\tau_{k}) W(\hat{\tau}_{k}) \rangle_{\Sigma_{1}}}\right)  \nonumber \\[+10pt]
&=Q_n+\f{\eta_{n}(A,M)}{n-1}\left( \f{n}{2} C^{A}_{VV}  C^{M}_{VV}   -C^{A}_{WW} \;C^{M}_{VV}  -\f{(n-2)}{2} C^{A}_{WW} C^{M}_{WW} \right) +\cdots,
\end{align} 
where
\beq
\eta_{n} (A,M) = \sum_{k=1}^{n-1} \langle J_{A}( \tau_{k}) J_{M}(\tau_{0}) \rangle_{\Sigma_n},
\eeq
In the appendices \ref{app:currentope} and \ref{app:analcont} we show that 
\bea
\lim_{n\rightarrow 1}\frac{1}{n-1}\eta_{n} (\tau ,\tau) &=  -C_J\f{\Gamma(\f{1}{2}) \Gamma(d)}{2^{2d-1}\Gamma( d+\f{1}{2})},\\
C^{M}_{VV} (\Sigma_1 : \tau-\hat \tau)&=i\frac{\rho_V}{C_J}\left( 2\sin \theta_{0} \right)^{d-1} \delta^{\tau M}  ,
\eea
where 
\beq
\rho_V=\langle V|J_0|V\rangle_{\mathbb{R}\times S^{d-1}},
\eeq
is the $U(1)$ charge density of the state $|V\rangle$ on the sphere and $C_J$ is the (positive) coefficient of the current two point function. Putting the pieces together we obtain
\be
S(\rho_{V}|| \rho_{W}) =\f{1}{4C_{J}}  (\rho_{V}-\rho_{W})^2 \f{\Gamma (\f{1}{2}) \Gamma(d)}{\Gamma(d+\f{1}{2})} (\sin \theta_{0})^{2d-2} +\cdots \label{eq:currentrel}.
\ee

\section{Holographic calculations} 
\label{section:holography}

The relative entropy between an excited state $|V \rangle$ and the ground state can be calculated holographically.  To see this it is useful to write the relative entropy in the following way
\be 
S(\rho_{V}||\rho_{0}) = \Delta \langle K_{{\rm vac}} \rangle -\Delta S, 
\ee
where  $K_{\rm vac}$ is the modular Hamiltonian of the vacuum reduced density matrix
\be
K_{{\rm vac}} = -\log \rho_{0},  \quad  \Delta \langle K_{{\rm vac}} \rangle ={\rm tr} [\rho_{V} K_{{\rm vac}}]-{\rm tr}[ \rho_{0} K_{{\rm vac}}]   ,
\ee
while
$\Delta S$ denotes the difference of the entanglement entropy between the two states, $\Delta S=S(\rho_V)-S(\rho_0)$. This latter is holographically computable by using the Ryu-Takayanagi formula
\be
S_{A} = \f{{\rm Area}(\gamma_{A})}{4G_{N}} ,
\ee
where $\gamma_{A}$ is the bulk minimal surface anchored to the boundary of  the subsystem $A$. This is enough to determine the relative entropy $S(\rho_{V}||\rho_{0})$ since the vacuum modular Hamiltonian $K_{\rm vac}$ can be expressed as a local integral of a stress tensor component in the case of ball shaped regions. In the following, we check our results \eqref{eq:relscalar}, \eqref{eq:cftrel} by replacing one of the excited states with the vacuum. The calculations relevant for these cases were done in \cite{Blanco:2013joa}, here we briefly summarize these and show that they agree with the CFT results  \eqref{eq:relscalar}, \eqref{eq:cftrel}. The strategy in \cite{Blanco:2013joa} is to consider either the AdS black brane (corresponding to the stress tensor contribution), the charged AdS black brane (corredponding to the current contribution) or a scalar condensate (corresponding to the scalar contribution) with the appropriate boundary expectation values of the dominant operators, and to work perturbatively in these. In this case, the holographic entanglement entropy expands as
\beq
S_A = S_0 + \varepsilon_V S_1+\varepsilon_V^2 S_2 +\cdots,
\eeq
where $\varepsilon_V$ is the boundary energy density. Linear contributions must be canceled with $\langle \Delta K\rangle$ due to the positivity of the relative entropy. As the vacuum modular Hamiltonian is linear in the boundary stress tensor this means that $\langle \Delta K\rangle$ is entirely canceled by the term $\varepsilon_V S_1$. We are then left with the relative entropy
\beq
S(\rho_V||\rho_0)=-(S_0-S(\rho_0))-\varepsilon_V^2 S_2 +\cdots.
\eeq
In the case of purely metric perturbations, $S_0=S(\rho_0)$ and the relative entropy is given by $-\varepsilon_V^2 S_2$ to quadratic order in $\varepsilon_V$. When we perturb with a scalar expectation value $\langle O\rangle$, we have 
\beq
S_0=S(\rho_0)+\langle O\rangle^2 S^{(2)}_0 +\cdots,
\eeq
and therefore, to leading order in $\langle O\rangle\equiv \langle V|O|V\rangle$ and $\varepsilon_V$ the relative entropy is given by $-\langle O\rangle^2 S^{(2)}_0$.
Note that by virtue of \eqref{eq:relscalar}, \eqref{eq:cftrel}, the perturbative treatment in $\varepsilon_V$ and $\langle O\rangle$ should give the exact leading contribution in the expansion in the subsystem size.

\subsection{Metric perturbations}

The gravity dual of an excited state with the
stress tensor expectation value
\be 
\langle V| T_{\mu \nu} | V \rangle = {\rm diag} \left( \varepsilon_{V},\; \f{\varepsilon_{V}}{d-1}, \;.\;. \;.\; \f{\varepsilon_{V}}{d-1} \right), 
\ee 
is the AdS black brane\footnote{Strictly speaking, the black brane is dual to an ensemble of high energy states but a single high energy state should have the same geometry outside of the horizon as this ensemble. Also, note that we are allowed to replace the dual AdS black hole by the planar one because we are only interested in the small subsystem limit.}. 
In the Fefferman Graham gauge, this reads as 
\be
ds^2 =\f{L^2}{z^{2}} \left(dz^2 + g_{\mu \nu}(z,x^{\mu}) dx^{\mu} dx^{\nu} \right),
\ee
where the spatial component of the metric takes the following form 
\be 
g_{ij}dx^{i}dx^{j} =T(z)\big( dr^2 +  r^2 d\Omega_{d-2}^2\big),
\ee
with
\bea
T(z)&=\left( 1+\frac{1}{4} \frac{z^d}{z_h^d}\right)^{\frac{4}{d}}, && z_h^d=\frac{d-1}{2} \frac{L^{d-1}}{\ell_P^{d-1}} \frac{1}{\varepsilon_V}.
\eea
Near the boundary $z=0$, $T(z)$ is expanded as follows
\be
T(z)= 1+ a \hat{\varepsilon}_{V} z^{d} -\f{d-4}{8} (a\hat{\varepsilon}_{V})^2 z^{2d} +\cdots, \quad  \hat{\varepsilon}_{V}=\f{\varepsilon_{V}}{d-1}, \quad a=\f{2}{d} \f{\ell_{P}^{d-1}}{L^{d-1}}
\ee
with $8\pi G_{N} =\ell_{P}^{d-2}$.
We need the holographic entanglement entropy $S_{A}$ when the subsystem $A$ is given by the ball shaped region with radius $R$
\be
A=\{(r, \Omega_{d-2}) \big| \; 0 <r<R \; \},
\ee
in the black brane background up to second order in the energy density $\varepsilon_{V}$
\be
S_{A} (\epsilon_{V})= S_{0} +\varepsilon_{V} S_{1} +\varepsilon_{V}^2 S_{2}+ \cdots.
\ee
This has been done in \cite{Blanco:2013joa}, here we briefly summarize the calculation.  
The bulk minimal surface profile is given by $(z(r), r, \Omega_{d-2})$, and the area functional is  
\be
A[z(r)]= \Omega_{d-2} L^{d-1}\int^{R}_{0}dr T(z(r))^{\f{d-2}{2}} r^{d-2}\s{ z'(r)^2+T(z(r))} \label{eq: areafunctional}.
\ee
When $\varepsilon_{V}=0$, i.e. in Poincar\'e AdS, the minimal surface is given by $z(r)= \s{R^2-r^2}$. By perturbatively solving the Euler Lagrange equation coming from \eqref{eq: areafunctional} for the extremal area and plugging the solution back into \eqref{eq: areafunctional}, we obtain\footnote{Note that the $\f{\ell_{P}^{d-1}}{L^{d-1}}$ part we obtained is the inverse of what is found in e.q. (3.55) of \cite{Blanco:2013joa}.}
\be
S_{2}=-\f{\pi^{3/2}d \Omega_{d-2}\Gamma(d-1)}{2^{d+1}(d+1) \Gamma(d+\f{3}{2})} \; \f{\ell_{P}^{d-1}}{L^{d-1}} R^{2d} \epsilon_{V}^2 \label{eq:finalhol}
\ee
In the theory dual to Einstein gravity, the coefficient of the stress tensor two point function $C_{T}$ is given by\footnote{See for example (3.10) of \cite{Perlmutter:2013gua} with $\lambda=0$, or (3.10) of \cite{Polishchuk:1999nh} for a derivation.} 
\be
C_{T}=\f{d+1}{d-1}\f{\Gamma(d+1)}{{\pi^{\frac{d}{2}}}\Gamma(\f{d}{2})}\f{L^{d-1}}{\ell_{P}^{d-1}}.
\ee
By substituting this into (\ref{eq:finalhol}) and using the identity 
\be
\Gamma(\f{d}{2}) \Gamma(\f{d-1}{2}) = 2^{2-d} {\sqrt{\pi}}\Gamma(d-1)
\ee
we recover the CFT result (\ref{eq:cftrel}).

\subsection{Scalar perturbations}

We can also recover \eqref{eq:relscalar} holographically by perturbing Poincar\'e AdS with a scalar condensate. This has also been done in \cite{Blanco:2013joa}, we only need to match the normalizations. A state with a boundary expectation value $\langle \tilde O \rangle$ for the scalar primary $\tilde O$ with dimension $\Delta$ corresponds to a bulk scalar $\phi$ with mass $m=\Delta(d-\Delta)$ and asymptotic profile\footnote{There are no sources turned on as we are interested in a CFT state.}
\beq
\label{eq:scalarasym}
\phi=\gamma \tilde O z^\Delta + \cdots,
\eeq
where the proper normalization has been determined in \cite{Klebanov:1999tb} and it is\footnote{Note that \cite{Klebanov:1999tb} uses $\eta=1$. To restore the $\eta$ dependence, note that (2.11) of \cite{Klebanov:1999tb} has the correct normalization for $\eta \neq 1$ but the on-shell action (2.16) has to be multiplied by $\eta$.}
\beq
\label{eq:normalise}
\gamma = \frac{1}{(2\Delta-d)\eta}.
\eeq
Here, $\eta$ is the coefficient of the kinetic term $-\frac{(\partial \phi)^2}{2}$ in the bulk action. The result of \cite{Blanco:2013joa} is that to quadratic order in the condensate the entanglement entropy changes as\footnote{Our final expression -- after the evaluation of the integral -- differs from (3.64) of \cite{Blanco:2013joa}.}
\bea
\Delta S &= -\frac{\pi \gamma^2L^{d-1}R}{4\ell_P^{d-1}}\langle\tilde O \rangle^2 \int_0^R \frac{dr \Omega_{d-2}r^{d-2}}{(R^2-r^2)^{\frac{d-2\Delta}{2}}} \left( 1-\frac{r^2}{(d-1)R^2}\right) \\
&=-\frac{\pi \gamma^2 L^{d-1} }{4 \ell_P^{d-1}}  \frac{\Delta \Gamma(\frac{d-1}{2})\Gamma(\Delta-\frac{d}{2}+1)}{2\Gamma(\Delta+\frac{3}{2})}\Omega_{d-2}\langle\tilde O \rangle^2 R^{2\Delta}.
\eea
Using the normalisation \eqref{eq:normalise} and the expression \eqref{eq:dsphere} for $\Omega_{d-2}$ we have
\beq
\label{eq:scalarpre}
\Delta S = -\frac{1}{\eta^2}\frac{L^{d-1}}{\ell_P^{d-1}} \frac{\Delta \pi^{\frac{d+1}{2}}\Gamma(\Delta-\frac{d}{2}+1)}{4(2\Delta-d)^2 \Gamma(\Delta+\frac{3}{2})}\langle\tilde O \rangle^2 R^{2\Delta}.
\eeq
The last step is to recall that via the standard AdS/CFT way of coupling $\tilde O$ to the asymptotic value of the bulk scalar $\phi$ we obtain the holographic two point function \cite{Klebanov:1999tb}
\beq
\label{eq:twopointnorm}
\langle \tilde O(x) \tilde O(0) \rangle =\eta (2\Delta-d)\frac{\Gamma(\Delta)}{\pi^{\frac{d}{2}} \Gamma(\Delta-\frac{d}{2})} \frac{1}{|x|^{2\Delta}},
\eeq 
 Our CFT calculations use normalized two point functions, therefore to match them we need to introduce $\tilde O=\sqrt{\eta (2\Delta-d)\frac{\Gamma(\Delta)}{\pi^{\frac{d}{2}} \Gamma(\Delta-\frac{d}{2})} }  O$. Plugging this into \eqref{eq:scalarpre} results in
\bea
\Delta S &= -\frac{1}{\eta} \frac{L^{d-1}}{\ell_P^{d-1}} \frac{\sqrt{\pi}\Gamma(\Delta+1)}{8\Gamma(\Delta+\frac{3}{2})}\langle O \rangle^2 R^{2\Delta}\\
&=
- \frac{\sqrt{\pi}\Gamma(\Delta+1)}{4\Gamma(\Delta+\frac{3}{2})}\langle O \rangle^2 R^{2\Delta},
\eea
where in the last line we have used that \cite{Blanco:2013joa} uses\footnote{We need to scale out $L^2$ from the $\sqrt{G}G^{\mu \nu}$ part from (3.56) of \cite{Blanco:2013joa} in order to match it with the action used in \cite{Klebanov:1999tb}. Note that both actions are dimensionless in natural units, so $\eta$ must be dimensionless as well.} $\eta=\frac{L^{d-1}}{2\ell_P^{d-1}}$. The result matches \eqref{eq:relscalar} as it should.

\subsection{Current perturbations}

To obtain the case when the dominant operator is a $U(1)$ current, one needs to calculate the holographic entanglement entropy in the charged black brane solution. This is again carried out in \cite{Blanco:2013joa}. Their result is
\beq
\Delta S = -\frac{\pi^{3/2}\Omega_{d-2}\Gamma(d-2)}{2^{d+1}\Gamma(d+\frac{1}{2})}\frac{L^{d-1}}{\ell_P^{d-1}} R^{2d-2}( \tilde J^0)^2
\eeq
There is one subtlety again. If we normalize the kinetic term of the bulk gauge field as $\frac{\eta}{4}F_{\mu\nu} F^{\mu  \nu}$ -- in \cite{Blanco:2013joa} $\eta=\frac{L^{d-1}}{2\ell_P^{d-1}}$, the same as for the scalar -- we have for the current one point function\footnote{This is deduced by comparing the bulk-to-boundary expression of the bulk gauge field, (47) of \cite{Freedman:1998tz}, to the one point function coming from the on-shell action formed from (54) of \cite{Freedman:1998tz}.} 
\beq
 J_0 = \eta \lim_{z\rightarrow 0} z^{3-d}\partial_z A_0,
\eeq
where $A_\mu$ is the bulk gauge field. Opposed to this, \cite{Blanco:2013joa} uses $\tilde J^0 = \lim_{z\rightarrow 0} z^{3-d}\partial_z A_0$. Therefore,
\beq
\Delta S = -\frac{\pi^{3/2}\Omega_{d-2}\Gamma(d-2)}{2^{d+1}\Gamma(d+\frac{1}{2})}\frac{4\ell_P^{d-1}}{L^{d-1}} R^{2d-2}(  J^0)^2.
\eeq
Now we can use the holographic expression for the two point function coefficient \cite{Freedman:1998tz}:
\beq
C_J =\frac{ (d-2)\Gamma(d)}{2 \pi^{\frac{d}{2}}\Gamma(\frac{d}{2})} \frac{L^{d-1}}{2\ell_P^{d-1}},
\eeq
to arrive at
\be
\Delta S =-\f{1}{4C_{J}}   \f{\Gamma (\f{1}{2}) \Gamma(d)}{\Gamma(d+\f{1}{2})} (J^0)^2 R^{2d-2} ,
\ee
in agreement with \eqref{eq:currentrel}.

\section{Beyond leading order in the subsystem size}
\label{sec:beyondsmall}

\subsection{The first asymmetric part of the relative entropy} 

In this section we discuss the first asymmetric part in the expansion of the relative entropy in the size of the subsystem by making use of the first law trick explained at the end of section \ref{subsection:firstlawtrick}.  For simplicity, we assume that the lightest operator $O$ is a scalar with conformal dimension $\Delta$ and that the conformal dimension $\Delta'$ of the next lightest operator satisfies $\Delta' > \f{3}{2} \Delta$.  Then, by a similar analysis as the one around (\ref{eq:trrhovn}), we obtain the following form for the entanglement entropy at third order in the $\theta_{0}$ expansion
\be
S(\rho_{W}) =S^{(2)}(\rho_{W}) +c^{(3)}_{\Delta}  \theta_{0}^{3\Delta}\la W | O |W \ra^{3} .
\ee
The coefficient $c^{(3)}_{\Delta}$ is given by the analytic continuation
\be
c^{(3)}_{\Delta}= \f{\p f^{(3)}}{ \p n} (\Delta,n) \big|_{n=1}, \quad   f^{(3)}(\Delta,n) = \sum_{(k,l,m)=0}^{n-1} \la O(2\pi(k-l)) O(2\pi(l-m))O(2\pi(m-k))\ra_{\Sigma_{n}}.
\ee
One can write   $c^{(3)}_{\Delta} = C_{OOO} b_{\Delta}$,
where $C_{OOO}$ denotes the three point function coefficient of $O$ while
 $b_{\Delta}$ is a universal constant that only depends on $\Delta$ and $d$. We were not able to perform the analytic continuation that is required to obtain $b_\Delta$ but we note that one can in fact compute it holographically and the result is found in e.q. (C.9) of \cite{Casini:2016rwj}. There, the third order change in the entanglement entropy due to perturbing with a scalar was calculated to be\footnote{We thank Damian Galante for his help regarding this formula.}
\beq
\delta S^{(3)} =- \pi 2 \eta \Omega_{d-2} R^{3\Delta} \phi^3 \frac{\kappa}{12} \frac{\Gamma(\frac{d-1}{2})\Gamma(\frac{3\Delta-d}{2})}{\Gamma(\frac{3\Delta+3}{2})},
\eeq
where the bulk cubic coupling is taken to be $\eta \frac{\kappa}{6} \phi^3$. 
We can use the holographic dictionary of \cite{Freedman:1998tz} to exchange $\kappa$ for the CFT OPE coefficient
\beq
\kappa = -\frac{1}{\eta} \frac{2\pi^d \Gamma(\Delta-\frac{d}{2})^3}{\Gamma(\Delta)^3\Gamma(\frac{3\Delta-d}{2})} C_{\tilde O \tilde O \tilde O},
\eeq
along with e.q.s (\ref{eq:scalarasym},\ref{eq:twopointnorm}) to write $\delta S^{(3)}$ in terms of CFT quantities
\beq
\delta S^{(3)} = \frac{2 \sqrt{\pi} }{3 \Gamma(\frac{3\Delta+3}{2})} \frac{1}{C_O^3} C_{OOO} \langle O\rangle^3 R^{3\Delta}.
\eeq
In this formula, we have allowed for an arbitrary two point function normalization $\langle O(x)O(0)\rangle = \frac{C_O}{|x|^{2\Delta}}$ for later convenience. Note the independence of the formula from the spacetime dimensionality $d$.

 Given this knowledge, we may write the next to leading order modular Hamiltonian as
\be
\label{eq:triopmodham}
K_{W} =K_{(2)}^{W} + K^{W}_{(3)}, \quad K^{W}_{(3)} =3c^{(3)}_{\Delta} \theta_{0}^{3\Delta}\la W | O |W \ra^{2} O, \quad c^{(3)}_{\Delta}=\frac{2 \sqrt{\pi} }{3 \Gamma(\frac{3\Delta+3}{2})} \frac{C_{OOO} }{C_O^3} ,
\ee
where $K^{W}_{(2)} $ is the leading order modular Hamiltonian of $| W \ra $ given in (\ref{eq:modex}). 
Then, the relative entropy reads as
\bea
\label{eq:triopcontrib}
S^{(3)} (\rho_{V} || \rho_{W})=S^{(2)}(\rho_{V} || \rho_{W}) - c^{(3)}_{\Delta}\theta_{0}^{3\Delta} (\la V |O |V\ra-\la W |O |W\ra)^2\left[ 2\la W |O |W\ra+\la V |O |V\ra \right],
\eea
where we have denoted the leading order result \eqref{eq:relent1} with $S^{(2)}(\rho_{V} || \rho_{W})$.

\subsection{An example in 2d}

In order to check the dependence on OPE coefficients in \eqref{eq:triopcontrib}, let us consider an excited state $|W \ra$ in a 2d CFT which has the following properties
\bea
\la W|T_{zz}| W \ra &=\la W|T_{\bar{z}\bar{z}}| W \ra= h_{V}, && \la W  |O | W \ra \approx 0,  \label{eq:classstate}
\eea
on the cylinder. Here, $O$ denotes an arbitrary primary operator different from the identity. We may think of $|W\rangle$ as some descendant of the vacuum or as a primary state for which all the three point function coefficients $C_{OWW}$ are sufficiently small\footnote{An example could be a state satisfying local ETH in the sense of \cite{Lashkari:2016vgj}.}. The modular Hamiltonian of such a state can be computed by pulling back the vacuum modular Hamiltonian on the plane\footnote{The coefficient of $K_{(w_{1}, w_{2})}$ is determined by 
\be 
K= 2 \pi\int \f{R^2-x^2}{2R} T_{00}dx
\ee
and our convention $T_{00}= -\frac{1}{2\pi} (T(z)+\bar{T}(\bar{z}))$. } 
\be
K_{(w_{1},w_{2})} =- \int^{w_{2}}_{w_{1}} \frac{(w_{2}-w)(w-w_{1})}{w_{1}-w_{2}}  T(w) dw, \label{eq:vacmod}
\ee
by the conformal map
 \bea
\label{eq:thermalmap}
w&= e^{\f{2\pi}{\beta_{W}}u}, && u&=\phi+i T,
\eea
where $\beta_W=\frac{2\pi}{\sqrt{24 h_W/c-1}}$. 
Note that this map is just the composition of two maps $w=z^{iT_{W}}$, $z=e^{-iu}$ where $T_W=\frac{1}{\beta_W}$. This map turns the vacuum expectation value of the stress tensor into its expectation value in the state $|W\rangle$. It is also just the map which brings the plane to the thermal cylinder with inverse temperature $\beta_W$. We choose the endpoints of the interval $A$ in \eqref{eq:vacmod} to be $w_{1}=1$ and $w_{2}=e^{i\ell}$. 
Applying the conformal map \eqref{eq:thermalmap} to the vacuum modular Hamiltonian gives
\be
K^{W}=  2 \beta_W\int^{\ell}_{0} \left(\f{\sinh \f{\pi(\ell-\phi)}{\beta_{W}}\sinh \f{\pi \phi}{\beta_{W}}}{\sinh \f{\pi \ell}{\beta_{W}}} \right) T_{00}(\phi) d\phi \label{eq:thermod}
\ee
This is just the expression for the modular Hamiltonian of a thermal state of temperature $T_W$ on the infinite line (see e.g. \cite{Hartman:2015apr}) but we can also interpret it as the modular Hamiltonian coming from the pure state $|W\rangle$ provided that $\ell \ll \beta_W$, i.e. we are in the thermodynamic limit. The reason for this requirement is the following. The periodicity of the cylinder $\phi \sim \phi +2\pi$ implies the identification $w \sim e^{\f{4\pi^2}{\beta_{W}}} w$. Therefore the image of the cylinder is not the entire plane, just the annulus  $\{ 1<|w|<e^{\f{4\pi^2}{\beta_{W}}} \}$. The modular Hamiltonian of the annulus vacuum is different from that of the plane vacuum $K_{(w_{1},w_{2})}$.  They agree only when the subsystem size $|w_{1}-w_{2}|$ is much smaller than the size of the annulus, i.e. 
\be
e^{\frac{2\pi}{\beta_W} \ell} \approx 1.
\ee
The entanglement entropy $S(\rho_{W})$ of the excited state \eqref{eq:classstate} can also be computed by applying the conformal map\footnote{Note that \eqref{eq:Hartman} can be derived directly from \eqref{eq:thermod}. Indeed, we may use the first law of entanglement $\la V|K^W|V\ra-\la W|K^W|W\ra \approx S(\rho_V)-S(\rho_W)$ for two states $|V\ra$ and $|W\ra$ with $h_V\approx h_W$. Taking the $h_V\rightarrow h_W$ limit yields
\beq
\nonumber
\frac{\partial S}{\partial h_W}=\frac{\beta_W}{\pi^2}\left( \pi \ell \coth\frac{\pi \ell}{\beta_W}-\beta_W\right),
\eeq
which integrates to \eqref{eq:Hartman} with constant of integration choosen to be $-\frac{c}{3}\log \epsilon$.} \eqref{eq:thermalmap} 
\beq
\label{eq:Hartman}
S(\rho_W)=\frac{c}{3} \log \left( \frac{\beta_W}{\pi \epsilon} \sinh \frac{\pi \ell}{\beta_W} \right),
\eeq 
which formally, of course, is just the entanglement entropy in the thermal state of an infinite line, see e.g. \cite{Calabrese:2009qy}. Now let $|V \ra$ be another excited state satisfying the condition \eqref{eq:classstate}. Then the relative entropy between $\rho_{W}$ and $\rho_{V}$ is given by\footnote{Again, formally this is the exact relative entropy between the reduced density matrices of two thermal states of temperatures $T_V$ and $T_W$ on the infinite line.}  
\bea
\label{eq:thermalrelative}
S(\rho_{V}|| \rho_{W}) &=\Delta \langle K_W \rangle - \Delta S \\
&=\frac{\beta_W}{\pi^2}\left( \pi \ell \coth\frac{\pi \ell}{\beta_W}-\beta_W\right)(h_V-h_W)-\frac{c}{3}\log \left( \frac{\beta_V \sinh \frac{\pi \ell}{\beta_V}}{\beta_W \sinh \frac{\pi \ell}{\beta_W}}\right).
\eea
Now let us expand this expression in powers of the interval size $\ell$
\beq
\label{eq:thermrelexpand}
S(\rho_{V}|| \rho_{W}) =\frac{1}{15c} (h_V-h_W)^2 \ell^4 + \frac{(h_V-h_W)^2(c-8(2h_W+h_V))}{315 c^2} \ell^6 + \cdots.
\eeq
We see that the leading behavior remarkably agrees with our previous computation in\footnote{With the conventions used there we need to make the replacement $\ell= 2\pi x$.} \cite{Sarosi:2016oks}, while in the subleading piece we can identify the trioperator contribution of the stress tensor, the same coefficient as in \eqref{eq:triopcontrib}. This is the piece that comes with $\frac{8}{315c^2}$ and we note that the holographic coefficient in \eqref{eq:triopmodham} exactly reproduces this factor if we sum up the cases when $O$ is the holomorphic or the antiholomorphic stress tensor. The part with $\frac{1}{315c}$ must be the bioperator contribution of higher descendants. 

To conclude this section, we note that as the expansion \eqref{eq:thermrelexpand} appears to be consistent with that of a pure state, even beyond leading order, it is tempting to consider the possibility that \eqref{eq:thermod} gives the vacuum contribution to the modular Hamiltonian of heavy states ($h_W \sim c$) in large $c$ CFTs up to $\frac{1}{c}$ corrections whenever the subsystem size does not exceed half of the total system, i.e. $\ell <\pi$. Intuitively, this is because one can create the BTZ spacetime from Poincar\'e AdS by applying a bulk coordinate transformation which asymptotes to \eqref{eq:thermalmap} (see e.g. \cite{Hubeny:2007xt}). We hope to return to this question and its possible applications in future work.

\section{Relative entropy and the variance of OPE coefficients}
\label{sec:variance}

To quantify how much states of a given energy are distinguishable in a given CFT it is useful to introduce quantities which do not depend on the particular choice of microstates. Such a quantity is the average relative entropy in a given energy window
\beq
\overline{S_{\rm rel}(E,\Delta E)}=\frac{1}{\rho^2}\sum_{i,j|E\leq \Delta_{V_i},\Delta_{V_{j}}\leq E+\Delta E} S(\rho_{V_i}||\rho_{V_j}),
\eeq
where $\rho=\rho(E,\Delta E)$ is the number of states in the energy window. Note that we are averaging positive quantities, therefore when this average is small, the individual relative entropies are guaranteed to be small. The purpose of this section is to point out that the result \eqref{eq:relent1} implies that in the small subsystem limit this average relative entropy is proportional to the microcanonical variance of the OPE coefficients in the given energy window, i.e.
\beq
\overline{S_{\rm rel}(E,\Delta E)}=\frac{\Gamma(\f{3}{2})\Gamma(\Delta+1)}{\Gamma(\Delta+\f{3}{2})}(\pi x)^{2\Delta}\sum_{\alpha} \overline{\Delta C^{O_\alpha}_{EE}}^2 + \cdots,
\eeq
where
\beq
\overline{\Delta C^{O_\alpha}_{EE}}^2 = \frac{1}{\rho}\sum_{i|E\leq \Delta_{V_i} \leq E+\Delta E} \left( C^{O_\alpha}_{V_iV_i}-\frac{1}{\rho}\left[\sum_{i|E\leq \Delta_{V_j} \leq E+\Delta E} C^{O_\alpha}_{V_jV_j}\right] \right)^2,
\eeq
is the variance and $\alpha$ runs over the set of lightest operators with dimensions $\Delta$ such that there is an operator in the energy window that has $O_\alpha$ in its OPE with itself. In holography this quantity is related to the distinguishability of bulk black hole microstates due to the fact that relative entropies are dual to each other \cite{Jafferis:2015del}. 

It is instructive to examine this quantity for the case when the lightest operator coupling to states of the given energy is the stress tensor itself. In this case we must use the formula \eqref{eq:higherdimstressrel} for the leading relative entropy, i.e. we have
\beq
\overline{S_{\rm rel}(E,\Delta E)} \sim \frac{1}{C_T} \overline{\Delta E^2}\theta_0^{2d},
\eeq
where $\overline{\Delta E^2}$ is the variance of the energy in the window $[E,E+\Delta E]$. A standard result on probability distributions with a compact support bounds this from above as\footnote{To prove this note that the variance is clearly maximal if all the probability is concentrated at the end points of the compact interval. In this case the variance is $\Delta E p(1-p)$ which is maximal at $p=\frac{1}{2}$.} $\overline{\Delta E^2} \leq \frac{(\Delta E)^2}{4}$. Now assume that we are interested in states with $E\sim N^2$ in large $N$ theories with $C_T \sim N^2$ and we set the window to be $\Delta E \sim E^\gamma \sim N^{2\gamma}$. In this case, the $N$ dependence of the average relative entropy is bounded as
\beq
\overline{S_{\rm rel}(E,\Delta E)} \lesssim N^{4\gamma-2},
\eeq
so that states become certainly indistinguishable in the large $N$ limit when the size of the window scales with exponent $\gamma < \frac{1}{2}$. This is well in line with the general expectation that black hole microstates are indistinguishable inside the microcanonical window\footnote{Note that for CFTs at large energies the variance of the energy in the canonical distribution scales with a higher power of $\bar E$ than $\frac{1}{2}$, namely $\overline{\Delta E^2} =\partial_\beta^2 \log Z \sim \bar E^{\frac{d+1}{d}}$. However, we are interested in the scaling in $N^2$ which is the extensive quantity. By standard saddle point reasoning, the width of the canonical distribution must scale in the thermodynamic limit as $\sqrt{M}$ if $M$ is an extensive quantity.}.

For general light operators, we note that the asymptotic $E\rightarrow \infty$ value of the microcanonical average part
\beq
\frac{1}{\rho}\left[\sum_{i|E\leq \Delta_{V_j} \leq E+\Delta E} C^{O_\alpha}_{V_jV_j}\right] 
\eeq
is fixed by modular invariance in two \cite{Kraus:2016nwo} and probably higher dimensions and it is (almost) universal. The missing piece\footnote{Some general bounds for this coming from crossing symmetry are available when $O_\alpha$ is much heavier than $V_j$ \cite{Pappadopulo:2012jk} and also there are universal large $c$ formulae for this quantity in the same regime \cite{Chang:2015qfa}. Unfortunately, this is the opposite of what we need here.} is the average of the square of $C^{O_\alpha}_{V_jV_j}$. It would be interesting to see whether 
\beq
\lim_{E\sim N^2,N\rightarrow \infty, \frac{\Delta E}{N} \rightarrow 0} \overline{\Delta C^{O_\alpha}_{EE}}^2=0,
\eeq
is satisfied by general conformal field theories, CFTs with a sparse low lying spectrum or is this a new requirement that a theory with weakly coupled gravity dual must satisfy. Finally, we note that this requirement appears to be strongly related to the assumption of local ETH of \cite{Lashkari:2016vgj}.





\section*{Acknowledgments}

We thank Alice Bernamonti, Pawel Caputa, Bartek Czech, Federico Galli, Stefan Leichenauer, Aitor Lewkowycz, M\'ark Mezei, Hiroshi Ooguri, Onkar Parrikar, and Bogdan Stoica for discussions. We thank Damian Galante for a helpful correspondence. Part of this work was explained in the YITP long term workshop "Quantum Information in String Theory and Many-body Systems".   
The work of T. U. was supported in part by the National Science Foundation under Grant
No. NSF PHY-25915. The work of G.S. was supported in part by a grant from the Simons Foundation (\#385592, Vijay Balasubramanian) through the It From Qubit Simons Collaboration, by the Belgian Federal Science Policy Office through the Interuniversity Attraction Pole P7/37, by FWO-Vlaanderen through projects G020714N and G044016N, and by Vrije Universiteit Brussel through the Strategic Research Program ``High-Energy Physics''.

\appendix 
\section{OPE coefficients}
\label{app:ope}
\subsection{Stress tensor contribution} 
\label{app:stressope}
In this section we compute the OPE coefficient in
\be
V (y_{1})V (y_{2}) =\langle V (y_{1})V (y_{2}) \rangle_{\mathcal{M}} \left[1+ \hat{C}^{MN}_{VV} (\mathcal{M}:y_{1}, y_{2}) \left( T_{MN} (y_{2})-\langle T_{MN}\rangle_\mathcal{M} \right) + \cdots \right].
\ee
The details of the function $ \hat{C}^{MN}_{VV} (\mathcal{M} : y_{1}-y_{2}) $ depend on the manifold $\mathcal{M}$ on which the CFT is defined. 
On flat space $\mathcal{M} =\mathbb{R}^{d}$ it is given by (see for example (c.5) of \cite{Komargodski:2016gci})
\be
\label{eq:flatstressope}
 \hat{C}^{\mu \nu}_{VV} (\mathbb{R}^{d} : s) =  \f{a}{C_{T}}\left[ \f{s^{\mu}s^{\nu}}{s^2}- \f{\delta_{\mu \nu}}{d} \right] s^{d}, \quad s^{\mu} =x_{1}^{\mu} -x_{2}^{\mu}.
\ee
The basic strategy to obtain this is to relate it to the flat space three point function $\langle T_{\mu\nu} V V\rangle_{\mathbb{R}^d}$ and then fix the normalization by mapping this latter to the matrix element $\langle V|T_{\mu\nu}| V\rangle_{\mathbb{R} \times S^{d-1}}$ on the cylinder by standard radial quantization. More explicitly, \eqref{eq:flatstressope} implies that in flat space we have \cite{Osborn:1993cr}
\beq
\langle V(\infty)T_{\mu\nu}(x)V(0)\rangle_{\mathbb{R}^d}=\frac{a}{x^d}\left( \frac{x_\mu x_\nu}{x^2}-\frac{1}{d}\delta_{\mu\nu}\right).
\eeq
Passing to polar coordinates $(r,\Omega_{d-1})$ and then by $e^{\tau}=r$ to $\mathbb{R}\times S^{d-1}$ we obtain
\beq
\langle V|T_{\mu\nu}(x)|V\rangle_{\mathbb{R}\times S^{d-1}} = a \; \text{diag}\left( \frac{d-1}{d},-\frac{1}{d},\cdots ,-\frac{1}{d}\right).
\eeq
The last step is to continue back to Lorentzian to obtain a proper matrix element. This amounts to an extra sign in $T_{00}$. The upshot is that $a$ is related to the energy density $\varepsilon_V$ as 
\be
a=-\f{d\; \varepsilon_{V}}{d-1}, \quad
\varepsilon_{V}=\f{\Delta_{V}}{\Omega_{d-1}}= \langle V|T_{00}| V\rangle_{\mathbb{R} \times S^{d-1}} ,  \quad \Omega_{d-1}= \f{2 \pi^{\f{d}{2}}}{\Gamma \left( \f{d}{2} \right)}. 
\ee
Here, $\Delta_V$ is the conformal dimension of the original operator $V$.
It is convenient to split  $\hat{C}^{\mu \nu}_{VV} (\mathbb{R}^{d} : s)$ into two parts
\be 
\hat{C}^{\mu \nu}_{VV} (\mathbb{R}^{d} : s) =C^{\mu \nu}_{\mathbb{R}^{d}}(s) +B(s)\; \delta^{\mu \nu}, \quad C^{\mu \nu}_{\mathbb{R}^{d}}(s)= \f{a}{C_{T}} \left( \f{s^{\mu}s^{\nu}}{s^2} \right)s^d, \;B_{\mathbb{R} }(s)=\f{a}{d C_{T}} s^{d}.
\ee
The term proportional to $\delta^{\mu \nu}$ corresponds to the exchange of the trace of the stress tensor $T^{\mu}_{\mu}$. Since flat space correlation functions involving $T^{\mu}_{\mu}$ contain only contact terms, we can ignore the contributions of them. When the displacement vector $s^{\mu}$ is purely timelike -- which will  be the  relevant  case for our later calculation --  only the time like component $C^{tt}_{\mathbb{R}^{d}}$ is nonvanishing.

When there is a conformal map between the manifold $\mathcal{M} $ and flat space one can derive the OPE coefficients $\hat{ C}^{MN}_{VV}(\mathcal{M}:)$ by mapping the two operators $V(y_{1}), V(y_{2})$  to flat space, using their OPE there and finally pulling the result back.
Let $\{y^{M}\}$ be the coordinates of $\mathcal{M} $ and $y^{M} (x^{\mu})$ be the conformal map with Weyl factor $\Omega$
\be 
ds^{2}_{\mathcal{M}}=\Omega^{2} ds^{2}_{\mathbb{R}^{d}}.
\ee
Then by using the transformation law of the stress tensor
\be
T_{\mu \nu}(x)=\Omega^{d-2}(x)\f{\p y^{M}}{\p x^{\mu}} \f{\p y^{N}}{\p x^{\nu}}\;\left( T_{MN} (y) {-\langle T_{MN} (y)  \rangle_\mathcal{M}}\right), \label{eq:transform}
\ee
we find the relation between OPE coefficients on flat space and $\mathcal{M}$
\begin{align} 
\hat{C}^{MN}_{VV} (\mathcal{M}: y_{1}-y_{2}) &=\Omega^{d-2} (y_{2}) \; \f{\p y^{M}}{\p x^{\mu}} \f{\p y^{N}}{\p x^{\nu}} C^{\mu \nu}_{VV} (\mathbb{R}^{d} :x_{1}-x_{2}) \label{eq:curvedope} \\[+10pt]
&\equiv C^{MN}_{\mathcal{M}} +B_{\mathcal{M}} g^{MN}.
\end{align}

Now we use this formula to compute the required OPE coefficients $C^{MN}_{S^{1} \times H^{d-1} } (y_{1}-y_{2})$. The manifold  $S^{1} \times H^{d-1}$ is conformally related to flat space $\mathbb{R}^{d}$ 
\be
ds^{2}_{S^{1} \times H^{d-1}} = \Omega^2 \left(  dt^2+ dr^2 +r^2 d\Omega_{d-2}^2 \right), \quad \Omega= \cosh u +\cos \tau
\ee
by the map 
\be
r= \f{\sinh u}{\cosh u +\cos \tau}, \qquad t= \f{\sin \tau}{\cosh u + \cos \tau}.  \label{eq:hyptoflat}
\ee
The relevant operator insertions in \eqref{eq:OPEstress} are at $y_{1}:(\tau= \pi- \theta_{0}, u=0)$ and $ y_{2}:(\tau= \pi+ \theta_{0}, u=0)$ on $ S^{1} \times H^{d-1}$. By using the map \eqref{eq:hyptoflat} these points are mapped to 
\be
y_{1} \rightarrow x_{1} : (t,r)= \left(\cot \f{\theta_{0}}{2}, 0 \right), \quad y_{2} \rightarrow x_{2} : (t,r)= \left(-\cot \f{\theta_{0}}{2}, 0 \right),
\ee
in flat space.
For this operator configuration the OPE coefficient  $C^{\mu \nu}_{\mathbb{R}^{d}}$ is purely time like
\be
C^{\mu \nu}_{\mathbb{R}^{d}}(x_{1}-x_{2})=\f{a}{C_{T}} \left(2 \cot \f{\theta_{0}}{2} \right)^{d} \delta^{\mu t}\delta^{\nu t}.
\ee
Using this flat space OPE coefficient and the Jacobian of the map (\ref{eq:hyptoflat})
\be
\f{\p \tau}{ \p t} \Big|_{r=0}=1+\cos \tau, \quad \f{\p u}{\p t} \Big|_{r=0} =0
\ee
we obtain $ C^{MN}_{S^{1} \times H^{d-1}}$ via the use of \eqref{eq:curvedope}
\be
C^{MN}_{S^{1} \times H^{d-1}} (y_{1}-y_{2})=\f{a}{ C_{T}}\left( 2\sin \theta_{0} \right)^{d} \delta^{\tau M}  \delta^{\tau N} \label{eq:opehyp}.
\ee

\subsection{Current contribution}
\label{app:currentope}

We proceed to compute the OPE coefficient in
\be
V (y_{1})V (y_{2}) =\langle V (y_{1})V (y_{2}) \rangle_{\mathcal{M}} \left[1+ \hat{C}^{M}_{VV} (\mathcal{M}:y_{1}, y_{2})  J_{M} (y_{2}) + \cdots \right],
\ee
for the case when the dominant contribution comes from a $U(1)$ current $J$, in an entirely analogous way as in the previous section. In flat space, the two point function and the OPE coefficient are fixed by conformal symmetry \cite{Osborn:1993cr}
\bea
\langle J_\mu(x)J_\nu(0)\rangle_{\mathbb{R}^d}&= \frac{C_J}{x^{2d-2}}\left(\delta_{\mu \nu}- \f{2 x^{\mu} x^{\nu}}{x^2} \right), \\
\hat{C}^{\mu}_{VV} (\mathbb{R}^d:s)&=\frac{b}{C_J} s^\mu s^{d-2},
\eea
where $C_J$ is the two point function normalization and $b$ is a coefficient not determined by conformal symmetry
. This latter is related to the flat space three point function by \cite{Osborn:1993cr}
\beq
\langle V(\infty)J_\mu(x)V(0)\rangle_{\mathbb{R}^d}=-b\frac{x_\mu}{x^d}.
\eeq
Passing to polar coordinates $(r,\Omega_{d-1})$ and then to $\mathbb{R}\times S^{d-1}$ by setting $e^\tau=r$ we obtain
\beq
\langle V|J_\mu(x)|V\rangle_{\mathbb{R}\times S^{d-1}} = \left( -b,0,\cdots,0 \right).
\eeq
Finally, we again continue to Lorentzian to obtain the matrix element which introduces an extra $i$ in the $J_0$ component. Therefore,
\beq
b=i\langle V|J_0|V\rangle_{\mathbb{R}\times S^{d-1}} \equiv i\frac{Q_V}{\Omega_{d-1}},
\eeq
i.e. the $U(1)$ charge density of $V$. Now we can calculate via the transformation rule
\beq
\label{eq:currentransf}
J_\mu(x)=\Omega^{d-2}(x) \f{\partial y^M}{\partial x^\mu}J_M(y),
\eeq
the OPE coefficient on $S^1\times H^{d-1}$ by applying the same conformal map as in 
the previous section \ref{app:stressope}. The result is
\beq
\hat C^{M}_{S^{1} \times H^{d-1}} (y_{1}-y_{2}) = \frac{b}{C_J}(2\sin \theta_0)^{d-1}  \delta^{M\tau}.
\eeq

\section{Analytic continuations}
\label{app:analcont}

In this appendix we review the required analytic continuations for both the scalar and the stress tensor contribution. The functions in the main text that we are required to continue to $n=1$ are
\bea
f(\Delta_O,n) &= \sum_{k=1}^{n-1} \la O( \tau_k) O (\tau_0)\ra_{S^{1}_{n} \times H^{d-1}},
\\
\eta_n(A,M) &= \sum_{k=1}^{n-1} \la J_A(\tau_k)J_M(\tau_0) \ra_{S^{1}_{n} \times H^{d-1}},\\
\sigma_{n} (AB,MN) &= \sum_{k=1}^{n-1} \langle T_{MN}( \tau_{k}) T_{AB}(\tau_{0}) \rangle_{S^{1}_{n} \times H^{d-1}},
\eea
see \eqref{eq: opeloc} for the definition of $\tau_k$. The
$n \rightarrow 1$ limits of these functions can be calculated by using the result of \cite{Agon:2015ftl} 
\bea
f(\Delta_O,n) &= (n-1) \int^{\infty}_{-\infty} \f{G^{OO}_{1}(-is+\pi)}{4 \cosh^2 \f{s}{2}} ds +O((n-1)^2), \\
\eta_{n} (\tau , \tau) &= (n-1) \int^{\infty}_{-\infty} \f{G^{JJ}_{1}(-is+\pi)}{4 \cosh^2 \f{s}{2}} ds +O((n-1)^2),\\
 \sigma_{n} (\tau \tau, \tau \tau) &= (n-1) \int^{\infty}_{-\infty} \f{G^{TT}_{1}(-is+\pi)}{4 \cosh^2 \f{s}{2}} ds +O((n-1)^2), \quad  \label{eq:sigmadef}
\eea
with
\bea
G^{OO}_{1}(s) &=\langle O(s)  O(0) \rangle_{S^{1} \times  H^{d-1}},\\
G^{JJ}_{1}(s) &=\langle J_{\tau}(s)  J_{\tau}(0) \rangle_{S^{1} \times  H^{d-1}}, \\
G^{TT}_{1}(s) &=\langle T_{\tau \tau}(s)  T_{\tau \tau}(0) \rangle_{S^{1} \times  H^{d-1}}, 
\eea
These two point functions are related to the flat space two point functions via the map \eqref{eq:hyptoflat}. The usual transformation rule is given by (\ref{eq:transform}) for the stress tensor, by \eqref{eq:currentransf} for the current and by
\beq
O(x)=\Omega(y)^{\Delta_O}O(y),
\eeq
for the scalar. 
In flat space the two point function of scalars and stress tensor elements are given by
\bea 
\langle O(x) O(0) \rangle_{\mathbb{R}^{d}} &= \frac{1}{x^{2\Delta_O}},\\
\langle J_\mu (x)J_\nu(0) \rangle_{\mathbb{R}^{d}}  &=\frac{C_J}{x^{2d-2}}I_{\mu\nu},\\
\langle T_{\mu \nu } (x) T_{\alpha \beta}(0) \rangle_{\mathbb{R}^{d}} &=\f{C_{T}}{x^{2d}}\; \mathcal{I}_{\mu \nu, \alpha\beta},
\eea
where 
\be
 \mathcal{I}_{\mu \nu, \alpha\beta} = \f{1}{2} \left( I_{\mu \alpha} I_{\nu \beta}+ I_{\mu \beta} I_{\nu \alpha} \right) -\f{\delta_{\mu\nu} \delta_{\alpha \beta}}{d}, \quad I_{\mu \nu} = \delta_{\mu \nu}- \f{2 x^{\mu} x^{\nu}}{x^2}
\ee
Therefore,
\bea
\langle O(s)  O(0) \rangle_{S^{1} \times  H^{d-1}} &=\f{1}{(2 \sin \frac{s}{2})^{2\Delta_O}},\\
\langle J_\tau(s)  J_\tau(0) \rangle_{S^{1} \times  H^{d-1}} &=-\frac{C_J}{\left[ 2\sin \frac{\theta_0}{2}\right]^{2d-2}},\\
\langle T_{\tau \tau}(s)  T_{\tau \tau}(0) \rangle_{S^{1} \times  H^{d-1}} &=\f{d-1}{d} \f{C_{T}}{\left[ 2\sin \f{s}{2}\right]^{2d}}.
\eea
The integral in (\ref{eq:sigmadef}) is calculated by 
\be
\int^{\infty}_{-\infty} \f{dx}{\cosh^{\nu}x } = B( \f{1}{2}, \f{\nu}{2}) =\f{\Gamma(\f{1}{2}) \Gamma(\f{\nu}{2})}{\Gamma(\f{1+\nu}{2})}
\ee
Using this in (\ref{eq:sigmadef}) we get 
\bea
f(\Delta_O,n) &=(n-1)\f{\Gamma(\f{1}{2}) \Gamma(\Delta+1)}{2^{2\Delta+1}\Gamma( \Delta+\f{3}{2})}+O((n-1)^2),\\
\eta_{n} (\tau, \tau) &= -(n-1)C_J\f{\Gamma(\f{1}{2}) \Gamma(d)}{2^{2d-1}\Gamma( d+\f{1}{2})}+O((n-1)^2)\\
 \sigma_{n} (\tau \tau, \tau \tau) &= (n-1) \f{C_{T}}{2^{2d+1}} \f{d-1}{d} \f{\Gamma (\f{1}{2}) \Gamma(d+1)}{\Gamma(d+\f{3}{2})}+O((n-1)^2). \label{eq:sigmastress}
\eea


\bibliographystyle{utphys}
\bibliography{higherdim}

\providecommand{\href}[2]{#2}\begingroup\raggedright\begin{thebibliography}{10}

\bibitem{Vedral:2002zz}
V.~Vedral, ``{The role of relative entropy in quantum information theory},''
\href{http://dx.doi.org/10.1103/RevModPhys.74.197}{{\em Rev. Mod. Phys.}
  {\bfseries 74} (2002) 197--234}.

\bibitem{ohya2004quantum}
M.~Ohya and D.~Petz, {\em Quantum entropy and its use}.
\newblock Springer Science \& Business Media, 2004.

\bibitem{Casini:2008cr}
H.~Casini, ``{Relative entropy and the Bekenstein bound},''
  \href{http://dx.doi.org/10.1088/0264-9381/25/20/205021}{{\em Class. Quant.
  Grav.} {\bfseries 25} (2008) 205021},
\href{http://arxiv.org/abs/0804.2182}{{\ttfamily arXiv:0804.2182 [hep-th]}}.

\bibitem{Bousso:2014sda}
R.~Bousso, H.~Casini, Z.~Fisher, and J.~Maldacena, ``{Proof of a Quantum Bousso
  Bound},'' \href{http://dx.doi.org/10.1103/PhysRevD.90.044002}{{\em Phys.
  Rev.} {\bfseries D90} no.~4, (2014) 044002},
\href{http://arxiv.org/abs/1404.5635}{{\ttfamily arXiv:1404.5635 [hep-th]}}.

\bibitem{Bousso:2014uxa}
R.~Bousso, H.~Casini, Z.~Fisher, and J.~Maldacena, ``{Entropy on a null surface
  for interacting quantum field theories and the Bousso bound},''
  \href{http://dx.doi.org/10.1103/PhysRevD.91.084030}{{\em Phys. Rev.}
  {\bfseries D91} no.~8, (2015) 084030},
\href{http://arxiv.org/abs/1406.4545}{{\ttfamily arXiv:1406.4545 [hep-th]}}.

\bibitem{Wall:2011hj}
A.~C. Wall, ``{A proof of the generalized second law for rapidly changing
  fields and arbitrary horizon slices},''
  \href{http://dx.doi.org/10.1103/PhysRevD.87.069904,
  10.1103/PhysRevD.85.104049}{{\em Phys. Rev.} {\bfseries D85} (2012) 104049},
  \href{http://arxiv.org/abs/1105.3445}{{\ttfamily arXiv:1105.3445 [gr-qc]}}.
[Erratum: Phys. Rev.D87,no.6,069904(2013)].

\bibitem{Faulkner:2016mzt}
T.~Faulkner, R.~G. Leigh, O.~Parrikar, and H.~Wang, ``{Modular Hamiltonians for
  Deformed Half-Spaces and the Averaged Null Energy Condition},''
  \href{http://dx.doi.org/10.1007/JHEP09(2016)038}{{\em JHEP} {\bfseries 09}
  (2016) 038},
\href{http://arxiv.org/abs/1605.08072}{{\ttfamily arXiv:1605.08072 [hep-th]}}.

\bibitem{Faulkner:2013ica}
T.~Faulkner, M.~Guica, T.~Hartman, R.~C. Myers, and M.~Van~Raamsdonk,
  ``{Gravitation from Entanglement in Holographic CFTs},''
  \href{http://dx.doi.org/10.1007/JHEP03(2014)051}{{\em JHEP} {\bfseries 03}
  (2014) 051},
\href{http://arxiv.org/abs/1312.7856}{{\ttfamily arXiv:1312.7856 [hep-th]}}.

\bibitem{Lin:2014hva}
J.~Lin, M.~Marcolli, H.~Ooguri, and B.~Stoica, ``{Locality of Gravitational
  Systems from Entanglement of Conformal Field Theories},''
  \href{http://dx.doi.org/10.1103/PhysRevLett.114.221601}{{\em Phys. Rev.
  Lett.} {\bfseries 114} (2015) 221601},
\href{http://arxiv.org/abs/1412.1879}{{\ttfamily arXiv:1412.1879 [hep-th]}}.

\bibitem{Lashkari:2014kda}
N.~Lashkari, C.~Rabideau, P.~Sabella-Garnier, and M.~Van~Raamsdonk,
  ``{Inviolable energy conditions from entanglement inequalities},''
  \href{http://dx.doi.org/10.1007/JHEP06(2015)067}{{\em JHEP} {\bfseries 06}
  (2015) 067},
\href{http://arxiv.org/abs/1412.3514}{{\ttfamily arXiv:1412.3514 [hep-th]}}.

\bibitem{Lashkari:2016idm}
N.~Lashkari, J.~Lin, H.~Ooguri, B.~Stoica, and M.~Van~Raamsdonk,
  ``{Gravitational Positive Energy Theorems from Information Inequalities},''
\href{http://arxiv.org/abs/1605.01075}{{\ttfamily arXiv:1605.01075 [hep-th]}}.

\bibitem{Lashkari:2014yva}
N.~Lashkari, ``{Relative Entropies in Conformal Field Theory},''
  \href{http://dx.doi.org/10.1103/PhysRevLett.113.051602}{{\em Phys. Rev.
  Lett.} {\bfseries 113} (2014) 051602},
\href{http://arxiv.org/abs/1404.3216}{{\ttfamily arXiv:1404.3216 [hep-th]}}.

\bibitem{Lashkari:2015dia}
N.~Lashkari, ``{Modular Hamiltonian for Excited States in Conformal Field
  Theory},'' \href{http://dx.doi.org/10.1103/PhysRevLett.117.041601}{{\em Phys.
  Rev. Lett.} {\bfseries 117} no.~4, (2016) 041601},
\href{http://arxiv.org/abs/1508.03506}{{\ttfamily arXiv:1508.03506 [hep-th]}}.

\bibitem{Sarosi:2016oks}
G.~S\'arosi and T.~Ugajin, ``{Relative entropy of excited states in two
  dimensional conformal field theories},''
  \href{http://dx.doi.org/10.1007/JHEP07(2016)114}{{\em JHEP} {\bfseries 07}
  (2016) 114},
\href{http://arxiv.org/abs/1603.03057}{{\ttfamily arXiv:1603.03057 [hep-th]}}.

\bibitem{Blanco:2013joa}
D.~D. Blanco, H.~Casini, L.-Y. Hung, and R.~C. Myers, ``{Relative Entropy and
  Holography},'' \href{http://dx.doi.org/10.1007/JHEP08(2013)060}{{\em JHEP}
  {\bfseries 08} (2013) 060},
\href{http://arxiv.org/abs/1305.3182}{{\ttfamily arXiv:1305.3182 [hep-th]}}.

\bibitem{Nakayama:2013is}
Y.~Nakayama, ``{Scale invariance vs conformal invariance},''
  \href{http://dx.doi.org/10.1016/j.physrep.2014.12.003}{{\em Phys. Rept.}
  {\bfseries 569} (2015) 1--93},
\href{http://arxiv.org/abs/1302.0884}{{\ttfamily arXiv:1302.0884 [hep-th]}}.

\bibitem{Karananas:2015ioa}
G.~K. Karananas and A.~Monin, ``{Weyl vs. Conformal},''
  \href{http://dx.doi.org/10.1016/j.physletb.2016.04.001}{{\em Phys. Lett.}
  {\bfseries B757} (2016) 257--260},
\href{http://arxiv.org/abs/1510.08042}{{\ttfamily arXiv:1510.08042 [hep-th]}}.

\bibitem{Casini:2011kv}
H.~Casini, M.~Huerta, and R.~C. Myers, ``{Towards a derivation of holographic
  entanglement entropy},''
  \href{http://dx.doi.org/10.1007/JHEP05(2011)036}{{\em JHEP} {\bfseries 05}
  (2011) 036},
\href{http://arxiv.org/abs/1102.0440}{{\ttfamily arXiv:1102.0440 [hep-th]}}.

\bibitem{Ryu:2006bv}
S.~Ryu and T.~Takayanagi, ``{Holographic derivation of entanglement entropy
  from AdS/CFT},'' \href{http://dx.doi.org/10.1103/PhysRevLett.96.181602}{{\em
  Phys. Rev. Lett.} {\bfseries 96} (2006) 181602},
\href{http://arxiv.org/abs/hep-th/0603001}{{\ttfamily arXiv:hep-th/0603001
  [hep-th]}}.

\bibitem{Ryu:2006ef}
S.~Ryu and T.~Takayanagi, ``{Aspects of Holographic Entanglement Entropy},''
  \href{http://dx.doi.org/10.1088/1126-6708/2006/08/045}{{\em JHEP} {\bfseries
  08} (2006) 045},
\href{http://arxiv.org/abs/hep-th/0605073}{{\ttfamily arXiv:hep-th/0605073
  [hep-th]}}.

\bibitem{nielsen2010quantum}
M.~A. Nielsen and I.~L. Chuang, {\em Quantum computation and quantum
  information}.
\newblock Cambridge university press, 2010.

\bibitem{Maldacena:1997re}
J.~M. Maldacena, ``{The Large N limit of superconformal field theories and
  supergravity},'' \href{http://dx.doi.org/10.1023/A:1026654312961}{{\em Int.
  J. Theor. Phys.} {\bfseries 38} (1999) 1113--1133},
  \href{http://arxiv.org/abs/hep-th/9711200}{{\ttfamily arXiv:hep-th/9711200
  [hep-th]}}.
[Adv. Theor. Math. Phys.2,231(1998)].

\bibitem{Witten:1998qj}
E.~Witten, ``{Anti-de Sitter space and holography},'' {\em Adv. Theor. Math.
  Phys.} {\bfseries 2} (1998) 253--291,
\href{http://arxiv.org/abs/hep-th/9802150}{{\ttfamily arXiv:hep-th/9802150
  [hep-th]}}.

\bibitem{Jafferis:2015del}
D.~L. Jafferis, A.~Lewkowycz, J.~Maldacena, and S.~J. Suh, ``{Relative entropy
  equals bulk relative entropy},''
  \href{http://dx.doi.org/10.1007/JHEP06(2016)004}{{\em JHEP} {\bfseries 06}
  (2016) 004},
\href{http://arxiv.org/abs/1512.06431}{{\ttfamily arXiv:1512.06431 [hep-th]}}.

\bibitem{Lin:2016dxa}
F.-L. Lin, H.~Wang, and J.-j. Zhang, ``{Thermality and excited state R\'enyi
  entropy in two-dimensional CFT},''
\href{http://arxiv.org/abs/1610.01362}{{\ttfamily arXiv:1610.01362 [hep-th]}}.

\bibitem{Lashkari:2016vgj}
N.~Lashkari, A.~Dymarsky, and H.~Liu, ``{Eigenstate Thermalization Hypothesis
  in Conformal Field Theory},''
\href{http://arxiv.org/abs/1610.00302}{{\ttfamily arXiv:1610.00302 [hep-th]}}.

\bibitem{Agon:2015ftl}
C.~Ag\'on and T.~Faulkner, ``{Quantum Corrections to Holographic Mutual
  Information},'' \href{http://dx.doi.org/10.1007/JHEP08(2016)118}{{\em JHEP}
  {\bfseries 08} (2016) 118},
\href{http://arxiv.org/abs/1511.07462}{{\ttfamily arXiv:1511.07462 [hep-th]}}.

\bibitem{Alcaraz:2011tn}
F.~C. Alcaraz, M.~I. Berganza, and G.~Sierra, ``{Entanglement of low-energy
  excitations in Conformal Field Theory},''
  \href{http://dx.doi.org/10.1103/PhysRevLett.106.201601}{{\em Phys. Rev.
  Lett.} {\bfseries 106} (2011) 201601},
\href{http://arxiv.org/abs/1101.2881}{{\ttfamily arXiv:1101.2881
  [cond-mat.stat-mech]}}.

\bibitem{DiFrancesco:1997nk}
P.~Di~Francesco, P.~Mathieu, and D.~Senechal,
  \href{http://dx.doi.org/10.1007/978-1-4612-2256-9}{{\em {Conformal Field
  Theory}}}.
\newblock Graduate Texts in Contemporary Physics. Springer-Verlag, New York,
  1997.
\newblock
\url{http://www-spires.fnal.gov/spires/find/books/www?cl=QC174.52.C66D5::1997}.
\newblock

\bibitem{Perlmutter:2013gua}
E.~Perlmutter, ``{A universal feature of CFT Rényi entropy},''
  \href{http://dx.doi.org/10.1007/JHEP03(2014)117}{{\em JHEP} {\bfseries 03}
  (2014) 117},
\href{http://arxiv.org/abs/1308.1083}{{\ttfamily arXiv:1308.1083 [hep-th]}}.

\bibitem{Polishchuk:1999nh}
A.~Polishchuk, ``{Massive symmetric tensor field on AdS},''
  \href{http://dx.doi.org/10.1088/1126-6708/1999/07/007}{{\em JHEP} {\bfseries
  07} (1999) 007},
\href{http://arxiv.org/abs/hep-th/9905048}{{\ttfamily arXiv:hep-th/9905048
  [hep-th]}}.

\bibitem{Klebanov:1999tb}
I.~R. Klebanov and E.~Witten, ``{AdS / CFT correspondence and symmetry
  breaking},'' \href{http://dx.doi.org/10.1016/S0550-3213(99)00387-9}{{\em
  Nucl. Phys.} {\bfseries B556} (1999) 89--114},
\href{http://arxiv.org/abs/hep-th/9905104}{{\ttfamily arXiv:hep-th/9905104
  [hep-th]}}.

\bibitem{Freedman:1998tz}
D.~Z. Freedman, S.~D. Mathur, A.~Matusis, and L.~Rastelli, ``{Correlation
  functions in the CFT(d) / AdS(d+1) correspondence},''
  \href{http://dx.doi.org/10.1016/S0550-3213(99)00053-X}{{\em Nucl. Phys.}
  {\bfseries B546} (1999) 96--118},
\href{http://arxiv.org/abs/hep-th/9804058}{{\ttfamily arXiv:hep-th/9804058
  [hep-th]}}.

\bibitem{Casini:2016rwj}
H.~Casini, D.~A. Galante, and R.~C. Myers, ``{Comments on Jacobson?s
  ?entanglement equilibrium and the Einstein equation?},''
  \href{http://dx.doi.org/10.1007/JHEP03(2016)194}{{\em JHEP} {\bfseries 03}
  (2016) 194},
\href{http://arxiv.org/abs/1601.00528}{{\ttfamily arXiv:1601.00528 [hep-th]}}.

\bibitem{Hartman:2015apr}
T.~Hartman and N.~Afkhami-Jeddi, ``{Speed Limits for Entanglement},''
\href{http://arxiv.org/abs/1512.02695}{{\ttfamily arXiv:1512.02695 [hep-th]}}.

\bibitem{Calabrese:2009qy}
P.~Calabrese and J.~Cardy, ``{Entanglement entropy and conformal field
  theory},'' \href{http://dx.doi.org/10.1088/1751-8113/42/50/504005}{{\em J.
  Phys.} {\bfseries A42} (2009) 504005},
\href{http://arxiv.org/abs/0905.4013}{{\ttfamily arXiv:0905.4013
  [cond-mat.stat-mech]}}.

\bibitem{Hubeny:2007xt}
V.~E. Hubeny, M.~Rangamani, and T.~Takayanagi, ``{A Covariant holographic
  entanglement entropy proposal},''
  \href{http://dx.doi.org/10.1088/1126-6708/2007/07/062}{{\em JHEP} {\bfseries
  07} (2007) 062},
\href{http://arxiv.org/abs/0705.0016}{{\ttfamily arXiv:0705.0016 [hep-th]}}.

\bibitem{Kraus:2016nwo}
P.~Kraus and A.~Maloney, ``{A Cardy Formula for Three-Point Coefficients: How
  the Black Hole Got its Spots},''
\href{http://arxiv.org/abs/1608.03284}{{\ttfamily arXiv:1608.03284 [hep-th]}}.

\bibitem{Pappadopulo:2012jk}
D.~Pappadopulo, S.~Rychkov, J.~Espin, and R.~Rattazzi, ``{OPE Convergence in
  Conformal Field Theory},''
  \href{http://dx.doi.org/10.1103/PhysRevD.86.105043}{{\em Phys. Rev.}
  {\bfseries D86} (2012) 105043},
\href{http://arxiv.org/abs/1208.6449}{{\ttfamily arXiv:1208.6449 [hep-th]}}.

\bibitem{Chang:2015qfa}
C.-M. Chang and Y.-H. Lin, ``{Bootstrapping 2D CFTs in the Semiclassical
  Limit},'' \href{http://dx.doi.org/10.1007/JHEP08(2016)056}{{\em JHEP}
  {\bfseries 08} (2016) 056},
\href{http://arxiv.org/abs/1510.02464}{{\ttfamily arXiv:1510.02464 [hep-th]}}.

\bibitem{Komargodski:2016gci}
Z.~Komargodski, M.~Kulaxizi, A.~Parnachev, and A.~Zhiboedov, ``{Conformal Field
  Theories and Deep Inelastic Scattering},''
\href{http://arxiv.org/abs/1601.05453}{{\ttfamily arXiv:1601.05453 [hep-th]}}.

\bibitem{Osborn:1993cr}
H.~Osborn and A.~C. Petkou, ``{Implications of conformal invariance in field
  theories for general dimensions},''
  \href{http://dx.doi.org/10.1006/aphy.1994.1045}{{\em Annals Phys.} {\bfseries
  231} (1994) 311--362},
\href{http://arxiv.org/abs/hep-th/9307010}{{\ttfamily arXiv:hep-th/9307010
  [hep-th]}}.

\end{thebibliography}\endgroup

\end{document}